\ifdefined\PRformat
\documentclass[usenatbib, twocolumn, nofootinbib, superscriptaddress, longbibliography, reprint, prl]{revtex4-1}
\else
\documentclass[usenatbib, twocolumn, twocolappendix, nofootinbib]{aastex701}
\fi

\usepackage{multirow}
\usepackage{hyperref}
\usepackage{bigints}
\usepackage{natbib}
\usepackage{color}
\usepackage{wrapfig}
\usepackage{amsmath,amsfonts,bm}%,amsfonts,amsthm,bm}
\usepackage{hyperref}% add hypertext capabilities
\usepackage[T1]{fontenc}
\usepackage{graphicx}   % Including figure files
\usepackage[mathlines]{lineno}
\usepackage{wrapfig}
\usepackage{fontawesome}
\usepackage{enumitem}
\usepackage{amsmath}

\linenumbers

\setlist[itemize]{align=parleft,left=0pt..1em}

\newcommand{\prsectiontitleformat}[1]{\textbf{\emph{#1}}. ---}
\ifdefined\PRformat
\else
\newcolumntype{C}[1]{>{\centering\let\newline\\\arraybackslash\hspace{0pt}}m{#1}}
\fi
\ifdefined\PRformat
\definecolor{brown}{rgb}{0.8, 0.5, 0.2}
\fi

\newcommand{\sptpol}{SPTpol}
\newcommand{\sptthreeg}{SPT-3G}

\newcommand{\act}{ACT}
\newcommand{\planck}{{\it Planck}}
\newcommand{\herschel}{{\it Herschel}}

\newcommand{\snr}{S/N}

\newcommand{\agora}{\textsc{Agora}}
\newcommand{\flamingo}{FLAMINGO}

\newcommand{\camb}{\texttt{CAMB}}

\newcommand{\fsky}{f_{\rm sky}}

\newcommand{\degree}{\ensuremath{^\circ}}
\newcommand{\pixres}{0.^{\prime}5}%0.^{\prime}5$}
\newcommand{\ukam}{\ensuremath{\mu}{\rm K{\text -}arcmin}}%^{\prime}}}
\newcommand{\uk}{\ensuremath{\mu}{\rm K}}%^{\prime}}}
\newcommand{\sqdeg}{{\rm deg}^{2}}
\newcommand{\mjy}{{\rm mJy}}

\newcommand{\msol}{\ensuremath{\mbox{M}_{\odot}}}
\newcommand{\munits}{\times 10^{14}~\msol}

\newcommand{\lcdm}{\Lambda{\rm CDM}}
\newcommand{\seight}{S_{8}}

\newcommand{\comment}[1]{}
\newcommand{\sz}{Sunyaev-Zel{'}dovich}

\newcommand{\mvir}{M_{\rm 500c}}

\newcommand{\ellmax}{\ell_{\rm max}}

\newcommand{\alpharad}{\alpha_{\rm rad}}
\newcommand{\alpharadsigma}{\sigma(\alpha_{\rm rad})}

\newcommand{\cibfree}{CIB-min}%minimized}
\newcommand{\radiofree}{Radio-min}%minimized}
\newcommand{\firstsptband}{95} %{90}

\newcommand{\clyy}{C_{\ell}^{yy}}
\newcommand{\clcov}{{\bf C}_{\rm \ell}}

\newcommand{\dlcmbres}{D_{\ell}^{yy, {\rm CMB-res}}}
\newcommand{\dlcibres}{D_{\ell}^{yy, {\rm CIB-res}}}
\newcommand{\dlkszres}{D_{\ell}^{yy, {\rm kSZ-res}}}
\newcommand{\dlradres}{D_{\ell}^{yy, {\rm Rad-res}}}
\newcommand{\rhotszcrosscib}{\rho_{\ell}^{\rm tSZ \times CIB}}
\newcommand{\tszcrosscib}{\mbox{tSZ $\times$ CIB}}

\newcommand{\mvcrosscibfree}{\mbox{MV $\times$ \cibfree}}
\newcommand{\lminforps}{500}
\newcommand{\lmaxforps}{5000}
\newcommand{\baselinemaskingthreshold}{2}
\newcommand{\howmanybundles}{100}

\newcommand{\howmanysimulations}{100}
\newcommand{\cibtweakingsigma}{0.2}
\newcommand{\howmanycibtweaks}{100}

\newcommand{\howmanyskypatches}{25}

\newcommand{\fieldsize}{\mbox{100 $\sqdeg$}}
\newcommand{\clusterstacksize}{6^{\prime} \times 6^{\prime}}

\newcommand{\cmbresidualcolour}{yellow}
\newcommand{\cibresidualcolour}{red}
\newcommand{\radioresidualcolour}{blue}
\newcommand{\kszresidualcolour}{green}
\newcommand{\tszpowerspectrumsnr}{9.3\sigma}
\newcommand{\tszpowerspectrumsnrformv}{9.4\sigma}
\newcommand{\tszpowerspectrumsnrforcibfree}{9.3\sigma}
\newcommand{\tszpowerspectrumsnrformvcrosscibfree}{9.4\sigma}
\newcommand{\snrofrhotszcibformvandcibfree}{3.1}
\newcommand{\snrofrhotszcibforcibfreeandmvcrosscibfree}{2}

\newcommand{\abstracttext}
{
We present a measurement of the full shape of the power spectrum of the thermal \sz{} (tSZ) effect down to arcminute scales using cosmic microwave background (CMB) data from the South Pole Telescope (SPT) over roughly \fieldsize{} field. 
The analysis incorporates data from the 2019/20 seasons of the SPT-3G survey in bands centered at \firstsptband, 150, and 220 GHz; from the full SPTpol dataset at 150 GHz; and from \herschel-SPIRE survey in bands centered at 600 and 857 GHz. 
We combine data from all the above bands using linear combination (LC) techniques to produce a tSZ or \mbox{Compton-$y$} map. 
We modify the LC weights to produce multiple versions of the Compton-$y$ map, including minimum-variance (MV) and foreground-minimized (-min) maps. 
We measure the auto- and cross-spectra of a subset of these maps in the range $\ell \in [\lminforps, \lmaxforps]$. 
While this power spectrum includes contributions from signals other than tSZ, we present numerous checks to show that the most challenging foreground signal, the cosmic infrared background (CIB) is much lower 
than the desired tSZ signal in the scales of interest in this work.
The final tSZ power spectrum is measured at $\tszpowerspectrumsnr$ with both the MV and \cibfree{} maps. 
Our results are consistent with those reported in other CMB surveys across the literature.
Using the difference in the tSZ power spectrum from MV and \cibfree{} maps, we reconstruct the scale‑dependent tSZ–CIB cross‑correlation $\rhotszcrosscib$, finding $\snrofrhotszcibformvandcibfree\sigma$ evidence for a nonzero correlation coefficient that is positive on large scales and approaches zero for $\ell > 2500$.
This result represents the deepest tSZ maps ever produced and provides new constraints that can help refine astrophysical feedback mechanisms and models of the intracluster medium.
}

\newcommand{\tittext}{Measurement of the Full Shape of the Thermal \sz{} Power Spectrum from  South Pole Telescope and \textbf{\emph{Herschel}}-SPIRE Observations}

\ifdefined\PRformat

\newcommand{\aap}{AAP}

\newcommand{\apjs}{ApJS}
\newcommand{\apjl}{ApJL}

\fi

\defcitealias{smith17}{SF17}
\defcitealias{ferraro18}{FS18}
\defcitealias{raghunathan23}{RO23}
\defcitealias{reichardt21}{R21}

\begin{document}

%\title{\textsc{\tittext}}
\title{\tittext}
\correspondingauthor{Srinivasan Raghunathan}\email{sriniraghuna@gmail.com}
\author[0000-0003-1405-378X]{S.~Raghunathan}
\affiliation{Department of Physics \& Astronomy, University of California, One Shields Avenue, Davis, CA 95616, USA}
\affiliation{Center for AstroPhysical Surveys, National Center for Supercomputing Applications, Urbana, IL, 61801, USA}
\email{sriniraghuna@gmail.com}

\author{P.~A.~R.~Ade}
\affiliation{School of Physics and Astronomy, Cardiff University, Cardiff CF24 3YB, United Kingdom}
\email{EMAIL}

\author{D.~Anbajagane}
\affiliation{Department of Astronomy and Astrophysics, University of Chicago, 5640 South Ellis Avenue, Chicago, IL, 60637, USA}
\affiliation{Kavli Institute for Cosmological Physics, University of Chicago, 5640 South Ellis Avenue, Chicago, IL, 60637, USA}
\email{dhayaa@uchicago.edu}

\author[0000-0002-4435-4623]{A.~J.~Anderson}
\affiliation{Fermi National Accelerator Laboratory, MS209, P.O. Box 500, Batavia, IL, 60510, USA}
\affiliation{Kavli Institute for Cosmological Physics, University of Chicago, 5640 South Ellis Avenue, Chicago, IL, 60637, USA}
\affiliation{Department of Astronomy and Astrophysics, University of Chicago, 5640 South Ellis Avenue, Chicago, IL, 60637, USA}
\email{adama@fnal.gov}

\author[]{B.~Ansarinejad}
\affiliation{School of Physics, University of Melbourne, Parkville, VIC 3010, Australia}
\email{behzad.ansarinejad@gmail.com}

\author[0000-0002-0517-9842]{M.~Archipley}
\affiliation{Department of Astronomy and Astrophysics, University of Chicago, 5640 South Ellis Avenue, Chicago, IL, 60637, USA}
\affiliation{Kavli Institute for Cosmological Physics, University of Chicago, 5640 South Ellis Avenue, Chicago, IL, 60637, USA}
\email{archipleym@uchicago.edu}

\author[]{J.~E.~Austermann}
\affiliation{NIST Quantum Devices Group, 325 Broadway Mailcode 817.03, Boulder, CO, 80305, USA}
\affiliation{Department of Physics, University of Colorado, Boulder, CO, 80309, USA}
\email{EMAIL}

\author[0000-0001-6899-1873]{L.~Balkenhol}
\affiliation{Sorbonne Universit\'e, CNRS, UMR 7095, Institut d'Astrophysique de Paris, 98 bis bd Arago, 75014 Paris, France}
\email{lennart.balkenhol@iap.fr}

\author[0000-0002-1623-5651]{D.~R.~Barron}
\affiliation{Department of Physics and Astronomy, University of New Mexico, Albuquerque, NM, 87131, USA}
\email{barrondarcy@gmail.com}

\author[0000-0001-9103-9354]{P.~S.~Barry}
\affiliation{School of Physics and Astronomy, Cardiff University, Cardiff, CF24 3AA, UK}
\email{BarryP2@cardiff.ac.uk}
 
% ...

\collaboration{all}{({\rm SPTpol} and SPT-3G Collaboration)}

\resetfootnotetrue 

\ifdefined\PRformat
\else
\shorttitle{{\rm tSZ} Power Spectrum from SPT}
\shortauthors{S. Raghunathan et al., SPT Collaboration}
%\shortauthors{\shortauthourlist}
\fi

%\collaboration{SPT-3G and SPTpol Collaboration} \noaffiliation

\begin{abstract}
\abstracttext{}
\end{abstract}

\ifdefined\PRformat
\maketitle
\fi
%%%%%%%%%%%%%%%%%%%%%%%%%%%%%%%%
%\section*{Responses to comments} \href{https://docs.google.com/document/d/1XTlnEBSKDFryrAgYV6MMcw1Rhvrhrnsdd4vW2896RLc/edit}{Link to google docs}
%\ifdefined\PRformat
%\\
%\fi

\ifdefined\PRformat
\prsectiontitleformat{Introduction}
\else
\section{Introduction}
\label{sec_introduction}
Cosmic microwave background (CMB) photons free-streaming towards us from the surface of last scattering interact with structures at late times to produce the secondary anisotropies of the CMB. 
Consequently, the secondary anisotropies encode crucial information about structure formation in the Universe and are known to be excellent probes of both cosmology and astrophysics. 
The most dominant secondary CMB anisotropies include gravitational lensing \citep{blanchard87, lewis06} and the \sz{} (SZ) effect \citep{sunyaev80, sunyaev80b}, which has been detected at high significance in CMB maps alone and in cross-correlations of CMB maps with other cosmological surveys \citep{benson04, jones05, smith07, staniszewski09, das11, bleem15b, omori17, planck18-8, madhavacheril24}. 
The SZ effect can be further decomposed into the thermal-SZ (tSZ) effect, which is the inverse Compton scattering of CMB photons off free electrons in ionized regions \citep{sunyaev80}, and the kinematic-SZ (kSZ) effect, which is the Doppler boosting of CMB photons due to the bulk velocity of free electrons \citep{sunyaev80b}. 
In this work, we focus on the tSZ effect. 

%Since the first detection of the tSZ effect at the location of galaxy clusters in the early 2000s \citep{benson04, jones05, bonamente06}, the tSZ effect has since emerged as one of the promising astrophysical and cosmological probes \citep{bocquet24}. 
Over the past 20 years, the tSZ effect has emerged as a promising astrophysical and cosmological probe \citep{komatsu02, battaglia12, planck15-24, bocquet24}.
The tSZ effect is now being routinely used by all major high-resolution CMB experiments to compile large samples of previously unknown galaxy clusters \citep{staniszewski09, bleem15b, planck15-24, huang20, bleem20, hilton21, bleem24, kornoelje25, archipley26}, including at high redshifts ($z \gtrsim 1.5$).
%, thanks to its redshift independent nature \citep{sunyaev80}. 
Meanwhile, the power spectrum of the tSZ signal is also an excellent probe of structure formation \citep[][for example]{komatsu02, hill13b, planck15-24, horowitz17, rotti21}. 
To take advantage of this, groups have produced maps of the tSZ signal using data from the Atacama Cosmology Telescope (\act, \citealt{madhavacheril20, coulton24}), the \planck{} satellite \citep{planck15-22, chandran23}, and the South Pole Telescope (SPT, \citealt{bleem22}). 
Among other applications, these maps have been used to probe the gas physics of the Universe through signatures of cosmological shocks at cluster boundaries \citep{baxter21, anbajagane22}, as well as through correlations of the tSZ signal with lensing and galaxy distributions \citep{hill14, sanchez23, mccarthy24}.

Recently, the tSZ effect has also been used extensively to study the impact of baryonic physics on structure formation, one of the leading candidates for the $\seight$ tension through small-scale suppression of the matter power spectrum \citep[][and references therein]{preston23}. 
Numerous groups have explored the impact of baryon physics on tSZ with the help of simulations \citep{moser22, hadzhiyska23, osato23, pandey23, elbers25a, lau25, mccarthy25}, while on the data side, work has mostly been limited to cross-correlations, for example, by cross-correlating tSZ maps with galaxy positions \citep{schaan21, sanchez23, das23, liu25, dalal25} and comparing them to hydrodynamical simulations with different feedback prescriptions \citep{mccarthy14, schaye23}. %with some additional evidence for the presence of feedback by also including a velocity-weighted stack of kSZ maps at the location of galaxy positions \citep{schaan21, hadzhiyska25, riedguachalla25} and lensing \citep{pandey25}.

The tSZ power spectrum is also sensitive to the 
%\refresponse{effects of baryonic feedback} \citep[][for example]{mccarthy14}. 
effects of baryonic feedback \citep[][for example]{mccarthy14}. 
Baryonic feedback pushes gas out of the haloes and introduces a tilt in the tSZ power spectrum at arcminute ($\ell \gtrsim 3000$) scales \citep{shaw10b, mccarthy14, elbers25a}. 
This effect could contribute to the strong suppression of the tSZ power spectrum reported by ACT \citep{dunkley13, louis25} and SPT \citep{george15, reichardt21} at $\ell = 3000$ compared to a simple extrapolation of the large-scale measurements reported by \planck{} \citep{planck13-21, bolliet18, tanimura22, chandran23}. 
However, simulations suggest that resolving this discrepancy through baryonic feedback requires a significant and unrealistic amount of feedback \citep{mccarthy14, mccarthy25}. 
Another possible source of the discrepancy is unmodeled systematics in the measured power spectra, most likely related to the cosmic infrared background (CIB) %\refresponse{and radio galaxies}, and this remains an open question.
and radio galaxies, and this remains an open question.
%The CIB is the diffusion signal component of the dusty star forming galaxies (DSFG) \citep{viero19} that tend to live within galaxy clusters filling in the tSZ decrements at bands $\nu \le {\rm 220\ GHz}$, observed by ACT and SPT. 
%The CIB is the diffuse signal component arising from dusty star-forming galaxies (DSFGs; \citealt{viero19}) which can partially fill in the tSZ decrements at frequencies $\nu \leq 220\ {\rm GHz}$---the most sensitive bands for ACT and SPT---owing to the fact that both signals trace the underlying large-scale structure \citep{erler18, butler22, kornoelje25}.
The CIB is the diffuse signal component arising from dusty star-forming galaxies (DSFGs; \citealt{viero19}).
%\refresponse{Both CIB and radio sources} 
Both CIB and radio sources can partially fill in the tSZ decrements at frequencies $\nu \leq 220\ {\rm GHz}$---the most sensitive bands for ACT and SPT---owing to the fact that both 
%\refresponse{the tSZ signal and one from the dusty/radio galaxies} 
the tSZ signal and one from the dusty/radio galaxies trace the underlying large-scale structure \citep{erler18, butler22, kornoelje25}.
The small-scale measurements of the tSZ and kSZ power spectra by ACT and SPT \citep{dunkley13, george15, reichardt21} have thus far relied on assuming templates normalized at $\ell = 3000$ for both the signal components (tSZ/kSZ) and the undesired ones like the CIB. 
Since the CIB power, which has been known to be hard to model at sub-mm bands \citep{viero19}, dominates the total power spectrum at $\ell \ge 3000$ \citep[see Fig. 2 of][hereafter \citetalias{reichardt21}]{reichardt21}, any mismatch in the assumed templates can result in significant bias in the inferred SZ results, as demonstrated by \citet[hereafter \citetalias{raghunathan23}]{raghunathan23}. 

In this work, we use data from SPT and \herschel-SPIRE to extract the full shape of the tSZ power spectrum in the multipole range $\ell \in [\lminforps, \lmaxforps]$, similar to the recent work by \citet{efstathiou25}, rather than a single $\ell = 3000$ data point as done in many of the previous works. 
Making use of the frequency dependence of different signal components, we combine the data from all frequency bands to derive the tSZ spectrum.
While the measured total power spectrum includes contributions from undesired signals, we remove our best guess of this contamination using simulations. 
As mentioned above, the CIB is one of the undesired components that is difficult to model, and we perform a number of checks to ensure that any residual CIB contribution is small. 
The other components ---CMB, kSZ\footnote{While the cross-correlation between kSZ and galaxies have been measured with high significance \citep{hill16, ferraro16, kusiak21}, the auto-power spectrum of kSZ has not yet been measured with high precision. Nevertheless, we show that varying its amplitude has minimal impact on our results.}, and radio--- are comparatively straightforward to model. 
We also include systematic error budgets that account for uncertainties in modeling these components. 
Consequently, our reported measurement uncertainties incorporate both statistical errors and these additional systematic contributions.
%We perform numerous checks to show that the residual level of this contamination, including the CIB, is much smaller compared to the expected level of the tSZ. The measurement errors include both the statistical errors and the systematic errors that arise due to uncertainty in the level of the undesired signals. 
We estimate tSZ-only bandpowers in the multipole range $\ell \in [\lminforps, \lmaxforps]$. 
We also study and quantify the impact of source and cluster masking on the tSZ power spectrum.

Throughout this work, we use the $D_{\ell}$ convention in plots. $D_\ell$ is related to the angular power spectrum $C_\ell$ through $D_{\ell} = \frac{\ell (\ell+1)}{2 \pi} C_{\ell}$.
\fi

%%%%%%%%%%%%%%%%%%%%%%%%%%%%%%%%
\ifdefined\PRformat
\prsectiontitleformat{CMB maps}
\else
%\section{CMB maps}
%\label{sec_cmb_maps}
\section{Datasets and methods}
\label{sec_datasets}
\fi
This work uses the same datasets as the kSZ trispectrum analysis by \citet{raghunathan24}.  
We summarize the datasets here and refer the reader to the appendix of \citet{raghunathan24} for more details. 
The datasets for this analysis come from \sptpol{} (150 GHz, \citealt{austermann12}), \sptthreeg{} (\firstsptband, 150, and 220 GHz, \citealt{benson14, bender18, sobrin22}), and \herschel-SPIRE (600 and 857 GHz, \citealt{pilbratt10, griffin10}) surveys of a roughly \fieldsize{} region centered at right ascension \mbox{(RA)=$352.5\degree$ (23h30m)} and declination (Dec.)=$-55\degree$. 
The \sptpol{} observations were carried out between 2012 and 2016 while the \sptthreeg{} observations are from the 2019 and 2020 seasons. 
The map depths at \firstsptband, 150, and 220 GHz correspond to 4.5, 3, and 16 \ukam{} respectively. 
To create the SPT maps, we filter the raw SPT time ordered data (TOD) and then bin them into flat sky maps in the Sanson-Flamsteed projection with a pixel resolution of $\pixres$ \citep{schaffer11}. 
The filtering and map-making procedures are similar to previous published SPT works \citep[for example]{schaffer11, dutcher21}. 
The filtering scheme is employed to remove excess noise along the scan direction $\ell_{x}$ and to reduce aliasing when the TOD are binned into maps. 
%\refresponse{
The $\ell_{x}$ filtering significantly removes our sensitivity to large-scale modes and hence for the rest of this analysis we only focus on modes $\ell > 500$. %}
%The individual frequency maps are then calibrated by cross-correlating with \planck. 
After the maps are made, we cross-correlate the individual frequency maps with \planck{} to obtain absolute calibration factors.

\subsection{Linear combination}
\label{sec_ilc}
We combine SPT and \herschel-SPIRE maps using the harmonic space linear combination (LC) technique \citep{cardoso08} to produce the final Compton-$y$ maps. 
Before doing so, we deconvolve the experiment-dependent filter transfer function and the frequency-dependent beam window function from all the bands. 
For SPT, the beam used in the deconvolution is the version appropriate for a 2.7 K blackbody spectrum, 
%and ignores any variation across different frequency bands 
and we refer to this as the ``CMB'' beam $B_{\ell}^{\rm CMB}$ (N. Huang et al., in preparation).
We also remove the signal from bright point sources detected in our maps. 
Specifically, we inpaint \citep{benoitlevy13, raghunathan19c} the location of both radio and dusty sources with flux above a certain threshold at 150 GHz ($S_{150}$) using the information from regions outside of the inpainting radius. 
The inpainting radius changes with the flux level of the sources and is $\sim 3^{\prime}$ on average. 
We follow the inpainting procedure, rather than masking, to avoid artifacts during the ILC step. 
Our baseline results are with the inpainting threshold of $S_{150} = \baselinemaskingthreshold\ \mjy$, chosen to significantly suppress the contamination from point source signals, but we also repeat the procedure for other cases such as $S_{150} \in [6, 10, 20]\ \mjy$ to quantify the change in tSZ power as a function of source masking. 
%Due to masking we lose roughly $\fskylost = 0.18$ for the baseline case with $S_{150} = \baselinemaskingthreshold\ \mjy$, and  $\fskylost = 0.07, 0.04, 0.02$ for $S_{150} = 6, 10, 20\ \mjy$ masking thresholds.
Due to masking, we lose roughly 18\% of the total map area for the baseline case with $S_{150} = \baselinemaskingthreshold\ \mjy$, and  7\%, 4\%, and 2\% for $S_{150} = 6, 10,$ and $20\ \mjy$ masking thresholds.

The covariance matrix for the LC can be decomposed into two constituents: (a) experimental noise and (b) contribution from undesired signals like CMB, CIB, kSZ and radio galaxies. 
The noise covariance is calculated by randomly splitting our data into two sets $M_{A}$ and $M_{B}$ and calculating the 2D power spectrum of the half-difference map $(M_{A} - M_{B})/2$.
We repeat the procedure 100 times and take the average power spectrum to reduce the scatter in the noise covariance matrix.
The covariance matrix of the undesired signals is calculated using the \agora{} simulations \citep{omori24}. 
We note that any mismatch between the covariance in the simulations and that of the real data can lead to sub‑optimal LC weights, but not to a bias in the recovered signal. 
However, in \citetalias{raghunathan23} we showed that the residual levels remain essentially unchanged when the \agora{}‑based covariance used to compute the LC weights is replaced with a covariance model derived from SPT measurements \citepalias{reichardt21}.

We produce four versions of the Compton-$y$ maps: minimum-variance (MV), CMB-nulled (CMB-free), CIB-minimized (\cibfree) and Radio-minimized (\radiofree), using the LC and constrained-LC (cLC, \citealt{remazeilles11}) techniques. 
More specifically, as described in \S\ref{sec_map_bundles_cross_spectra}, we combine our observations into multiple map bundles and produce four versions of the Compton-$y$ maps for each bundle. 
This bundling-approach is to avoid the noise bias when calculating the power spectrum and more details are provided in \S\ref{sec_map_bundles_cross_spectra}. 
The cLC approach allows minimizing the contribution from particular components.
In cLC, the weights for each channel change compared to MV-LC based on the component being nulled or minimized.
For this, we assume a spectral energy distribution (SED) to either remove or minimize CMB, CIB, and radio contamination \citepalias[see Eq.(2) of][]{raghunathan23}. 
We use the differential CMB temperature units $\Delta T_{\rm CMB} = \Delta I_{\nu} \left[ \frac{\partial B_{\nu}(T, \nu)}{\partial T} \left. \right\vert_{T_{\rm CMB}} \right]^{-1}$ with respect to the average $T_{\rm CMB} = 2.7255 {\rm K}$ where $\nu$ is the observed frequency \citep{planck13-9}.
In these units, the SED for CMB is unity in all frequency bands. 
We model the frequency dependence of the radio sources using a power law with $\Delta I_{\nu} \propto \nu^{\alpha_{\rm radio}}$ with $\alpha_{\rm radio} = -0.76$ \citepalias{reichardt21}. 
For CIB, we use a modified blackbody formalism of the form $\Delta I_{\nu} \propto \nu^{\beta_{\rm CIB}} B_{\nu}(T_{\rm CIB})$ where $T_{\rm CIB}$ is the temperature, $\beta_{\rm CIB}$ is the emissivity index, and $B_{\nu}(T)$ is the \planck{} function.
%The SED for CMB is unity in all the frequency bands while for radio, we use a spectral index of $\alpha_{\rm radio} = -0.76$ \citepalias{reichardt21}. For CIB, we use a modified blackbody formalism of the form $\eta_{\nu} = \nu^{\beta_{\rm CIB}} B_{\nu}(T_{\rm CIB})$ where $T_{\rm CIB}$ is the temperature, $\beta_{\rm CIB}$ is the emissivity index, and $B_{\nu}(T)$ is the \planck{} function. 
%We use the grid search approach described in \citetalias{raghunathan23} to get a value of $T_{\rm CIB}$ and $\beta_{\rm CIB}$ that leads to the lowest level of CIB contamination in \agora{} simulations at $\ell \in [\lminforps, \lmaxforps]$ given the noise in each of the frequency bands. 
We use the grid‑search procedure described in \citetalias{raghunathan23} to determine the values of $T_{\rm CIB}$ and $\beta_{\rm CIB}$ that minimize the level of CIB contamination in the \agora{} simulations over $\ell \in [\lminforps, \lmaxforps]$, given the noise properties of each frequency band. 
It is important to note that this method yields the lowest possible CIB residual for the chosen set of frequency bands and their associated noise levels; however, it does not guarantee that the resulting CIB residual is smaller than the tSZ signal itself.
To account for this, as well as the fact that the SED used for the \cibfree{} map is based on \agora, we perform numerous checks in \S\ref{sec_cib_residuals} to ensure that any residual CIB contamination is much smaller than the desired tSZ signal. 
%\refresponse{
Another possible strategy is to use the moment-based CIB nulling technique, which removes the contribution from a chosen CIB SED as well as its first derivative, following \citet{chluba17} and as demonstrated in \citet{madhavacheril20, coulton24}. However, this method introduces a noise penalty and increases CMB residuals, both of which lead to a substantial growth in the final uncertainties \citep[see Fig.~15 of][]{coulton24}. For these reasons, we do not adopt this approach here. %}

\begin{figure}
\centering
\includegraphics[width=0.48\textwidth, keepaspectratio]{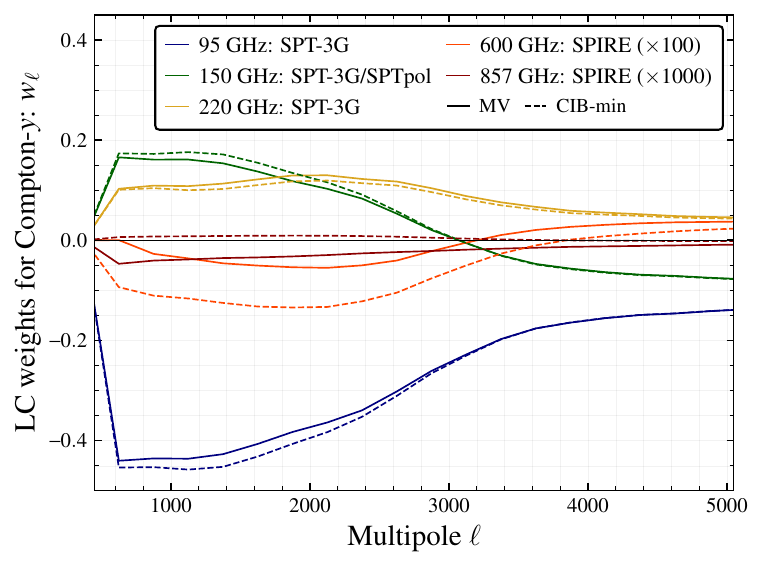}
\caption{Azimuthally averaged LC weights for the MV and the \cibfree{} Compton-$y$ maps.
The solid (dashed) curves correspond to MV (\cibfree) maps. 
Since the weights are much smaller for the \herschel-SPIRE 600 (857) bands compared to the other bands, they have been multiplied by $\times 100$ ($\times 1000$) for clarity in this plot. 
As mentioned in the text, for the ease of analysis, we assume same weights for the 150 GHz channel from both \sptthreeg{} and \sptpol.
%The weights for the 150 GHz for \sptthreeg{} and \sptpol, shown as green shades, are nearly identical, reflecting the similar noise levels and beam profiles of these bands.
}
\label{fig_lc_weights}
\end{figure}

Although the 150 GHz channels from \sptthreeg{} and SPTpol have slightly different noise levels and beams, we assume identical LC weights for both channels to simplify the analysis. 
However, we deconvolve the appropriate beam $B_{\ell}^{\rm CMB}$ from each of them.
%Srini: note for later: The tSZ frequency dependence for SPT-3G 150 (with bandpass) vs SPTpol 150 (w/o bandpass) is ~1 per cent.
In Fig.~\ref{fig_lc_weights}, we show the LC weights for MV and \cibfree{} maps as solid and dashed curves, respectively. 
%\refresponse{
As noted earlier, the scan‑direction filtering in our maps removes large‑scale modes $\ell_{x} \le 500$ and $\ell \le 500$, so the weights are strongly filtered on these scales and approach zero. %}

The CMB-nulled and \radiofree{} maps, because of the relative weighting of the bands required to reduce the contribution from CMB and radio,  are noisier than the others, and using simulations, we find that the residual CIB contribution in them is significantly higher than the expected level of tSZ on all scales. 
Consequently, we do not use them further in this analysis but they are available to be downloaded as part of the associated data release. 

\subsection{Filter transfer function and beams}
\label{sec_tf_beams}
We capture the effect of filtering by calculating the transfer function (TF) using mock-observations of Gaussian simulations. 
The TF is the average ratio of the power spectrum of 250 mock-observed maps to the input spectrum used to generate the simulations. 
We calculate the TF in 2D as a function of both $\ell_{x}$ and $\ell_{y}$ to capture the anisotropic nature of the filtering. 
For \herschel-SPIRE, to remove excess noise on large scales, we high-pass filter the data at $\left| \ell_{x} \right| \lesssim 200$ and $\left| \ell_{y} \right| \lesssim 200$. 
The frequency-dependent beam window functions $B_{\ell}^{\nu}$ for SPT are estimated using a combination of dedicated observations of planets and point sources in the CMB field data (\citealt{dutcher21}, Huang et al., in preparation). 
%For reference, the SPT beams in the \firstsptband, 150, and 220 GHz bands roughly correspond to $\theta_{\rm FWHM} = 1.^{\prime}7$, $1.^{\prime}2$, $1^{\prime}$. 
For reference, the SPT beams in the \firstsptband, 150, and 220 GHz bands can be approximated as Gaussians of width $\theta_{\rm FWHM} = 1.^{\prime}6$, $1.^{\prime}2$, $1^{\prime}$.
We use Gaussian beam approximations for \herschel-SPIRE with $\theta_{\rm FWHM} = 36.^{\prime \prime}6$ and $25.^{\prime \prime}2$ for the 600 and 857 GHz bands \citep{viero19}.

The variation of the experimental beam for different SEDs could be an important factor in the presence of foreground signals \citep{giardiello25}. 
We take this into account by modeling the main lobe of the \sptthreeg{} beam for different SEDs to produce beams for multiple components: $B_{\ell}^{\rm CIB}$, $B_{\ell}^{\rm Radio}$, and $B_{\ell}^{\rm tSZ}$ from an ongoing work (N. Huang et al., in preparation). 
The sidelobes are assumed to be frequency-independent.
These component-dependent beams differ from the $B_{\ell}^{\rm CMB}$ by: $\lesssim 2\%$ for $B_{\ell}^{\rm Radio}$; $\lesssim 5\%$ for $B_{\ell}^{\rm tSZ}$; and $\lesssim 10\%$ for $B_{\ell}^{\rm CIB}$ in the desired range of scales $\ell \in [\lminforps, \lmaxforps]$. 
The variation is more for 220 GHz compared to \firstsptband/150 GHz. 

\subsection{Map bundles and cross-spectra}
\label{sec_map_bundles_cross_spectra}
We follow previous works (\citetalias{reichardt21}, \citealt{dutcher21, louis25, camphuis25}) and calculate the tSZ power spectrum as the average of cross-power spectra between multiple Compton-$y$ map bundles. 
Since the noise is different in each bundle, the cross spectra will not, on average, contain a contribution from the experimental noise. 
This ensures that we are insensitive to the mis-estimation of the noise power spectrum, which, in an auto-spectrum analysis, could lead to biases in the final tSZ power spectrum \citep{polenta05, tristram05}. 
%Note, however, that we do use the power spectrum of the half-difference maps, as explained above, for the noise covariance in the LC step. The mis-estimation of the noise power in the LC step will only result in sub-optimal weighting of different frequency bands and not a bias. 

To this end, we randomly divide the SPT observations into \howmanybundles{} different bundles and take the inverse-variance weighted average of all the observations in each bundle. 
The bundles are chosen such that the noise is roughly the same across all the different bundles. 
On the other hand, we only have two bundled maps $M_{A}^{\rm SPIRE}$ and $M_{B}^{\rm SPIRE}$, rather than \howmanybundles, for the \herschel-SPIRE 600/857 GHz bands.  Consequently, we pick $M_{A}^{\rm SPIRE}$ ($M_{B}^{\rm SPIRE}$) for all the even (odd) bundles in the LC step to create the Compton-$y$ map. 
Before computing the power spectrum, we mask the inpainted locations in the Compton-$y$ maps since the locations do not contain the true tSZ signal. 
Using simulations, we estimate that the mode-coupling that might arise due to masking is negligible and ignore it in the analysis. 
We compute the final binned tSZ power spectrum $\hat{C}_{b}^{yy}$ in the range $\ell \in [\lminforps, \lmaxforps]$ in bins $b$ of width $\Delta \ell = 500$ as:
\begin{equation}
\label{eq_cross_spectra}
\hat{C}_{b}^{yy} = \frac{1}{\fsky} \frac{1}{N}\sum_{\substack{i {\rm = odd} \\ j {\rm = even}}}\ (1-\delta_{ij}) \sum_{\ell \in b} \frac{m_{\ell_{i}} m_{\ell_{j}}^{\ast}}{\Delta \ell}
%%%%Good%\hat{C}_{b, ij}^{yy} = \frac{1}{\fsky} \frac{1}{N}\sum_{\substack{i \% 2 = 0 \\ j \% 2 = 1}\ {\rm or}\ i \neq j}\ \sum_{\ell \in b} \frac{m_{\ell_{i}} m_{\ell_{j}}^{\ast}}{\Delta \ell}
%old%\hat{C}_{b, ij}^{yy} = \frac{1}{\fskyfinal} \frac{1}{N}\sum_{i \neq j}\ \sum_{\ell \in b} \frac{m_{\ell_{i}} m_{\ell_{j}}^{\ast}}{\Delta \ell}
\end{equation}
%where $\fskyfinal = 1 - \fskylost$ is the unmasked sky fraction, 
where $\fsky$ is the unmasked sky fraction, $N$ is the total number of cross-spectra, $m_{\ell}$ is the Fourier transform of the Compton-$y$ map $m$ and the conditions %$i \% 2 = 0$ and $j \% 2 = 1$ 
$(i = {\rm odd})$ and $(j = {\rm even})$ 
ensure that we do not compute the cross-spectra of pairs of even bundles or pairs of odd bundles since such pairs contain the same \herschel-SPIRE 600/857 GHz data. 
%The condition $i \neq j$ 
The condition $(1- \delta_{ij})$ 
ensures that we do not compute the auto‑spectrum of any bundle, as it would be affected by the noise bias of that same bundle.
%We note that relaxing the even/odd condition during the power spectra calculation has negligible impact on our results since the noise in \herschel-SPIRE maps are much smaller than the signal on all scales of interest \srini{check this}. 
%We note that relaxing the even/odd condition during the power spectrum calculation has negligible impact on our results since the \herschel-SPIRE maps are dominated by signal (CIB) on all scales of interest, with $\snr \sim 3 (8)$ at $\ell <3000$ ($\ell \ge 3000$) for both the \herschel-SPIRE channels. However, note that we never compute the auto-spectrum of any bundle and always set $ i \neq j$ even when relaxing the condition for even/odd bundles. 
In the subsequent sections we refer to the binned power spectrum $\hat{C}_{b}^{yy}$ simply as $C_{\ell}^{yy}$. 

\subsection{Simulations}
\label{sec_sims}
%Following \citet{raghunathan24}, we use two sets of simulations in this analysis: (a) \agora{} simulations of the individual frequency bands and (b) \howmanygaussiansimulations{} Gaussian realizations of the \agora{} power spectra. In both cases, we also include experimental noise realizations either using one of the half-difference maps for SPT or using Gaussian realizations of the noise spectra for \herschel-SPIRE. 
We use the \agora{} \citep{omori24} extragalactic simulations and include experimental noise realizations either using one of the half-difference maps for SPT or using Gaussian realizations of the noise spectra for \herschel-SPIRE.
We perform the exact same operations as done to the actual data before combining the simulations of the individual frequency bands using the LC step. 
These steps include filtering, convolving the simulations with the experimental beam, and inpainting. 
Note that we apply the component-dependent beams described in \S\ref{sec_tf_beams} to each of the foreground components in the simulations to properly mimic the real observations.
The locations of the inpainted sources are different in simulations compared to data and are determined by combining the CIB and radio portions of \agora{} at 150 GHz. 
We pick the locations of pixels above the masking threshold in the combined simulation and determine the masking radius based on the signal-to-noise ($\snr$) of sources with equivalent flux in data.

The \agora{} simulation suite contains correlated multi-component sky signals, namely CIB, CMB, late-time kSZ, radio and tSZ. 
Since the correlation between CIB, radio, and tSZ can be important for SZ inference (\citealt{dunkley13}, \citetalias{reichardt21}), and since we do not have robust high-$\snr$ measurements of the cross-correlations, we cannot fully trust the CIB/radio/tSZ correlations in \agora.
%Subsequently, sometimes we randomly swap the location of the patch of each of these components to explicitly break the correlations between components in \agora. 
Subsequently, we also randomly swap the location of the patch of each of these components to explicitly break the correlations between components in \agora. 
We then inject different levels of scale-dependent correlations, as discussed in \S\ref{sec_tsz_cib}, to check the importance of these correlations.
When doing so, we prefix each of the swapped components with \texttt{Uncorr-}.
Finally, \agora{} does not contain the contribution from the early-time reionization kSZ, and to include that, we generate Gaussian realizations with an underlying power spectrum corresponding to \mbox{$C_{\ell}^{\rm kSZ, reion} = \dfrac{1.5 (2 \pi)}{\ell (\ell+1)}\ \uk^{2}$}. 
Varying the level of reionization kSZ has no impact on our results. 

The simulations are used for the following: computing the LC weights; determining  the SED for CIB-min; obtaining the bandpower covariance matrix; performing systematic checks; and finally for estimating and removing the contribution of undesired signals (CIB, CMB, kSZ and radio) from the total measured spectrum to get the tSZ-only power spectrum. 
Since we only have a single realization of the \agora{} simulation, we extract \howmanysimulations{} non-overlapping \fieldsize{} patches corresponding to the size of our field \citep{raghunathan24}.  %To ensure that the patches can be treated as independent realizations, we perform a test by comparing the diagonal portion of the bandpower covariance using aversion computed from \howmanygaussiansimulations{} Gaussian realizations. \srini{This is from an old result where the clusters are masked. I will either confirm that old result or remove the references to the Gaussian sims.} If the patches are overlapping, we would expect a lower scatter in the bandpower covariance from \agora{} compared to Gaussian realizations. But we find an excellent agreement between the two sets of covariance matrices. Besides this check, the Gaussian realizations are not used for any other part of this analysis. For the rest of the analysis, we use the bandpower covariance matrix from \agora{} simulations since they capture the off-diagonal components of the covariance matrix due to the non-Gaussian nature of the tSZ and the point sources. 

%%%%%%%%%%%%%%%%%%%%%%%%%%%%%%%%
\section{Results}
\label{sec_results}

\subsection{Compton-$y$ maps and power spectrum measurement}
\label{sec_ymaps_power_spectrum}

\begin{figure}
\centering
\includegraphics[width=0.48\textwidth, keepaspectratio]{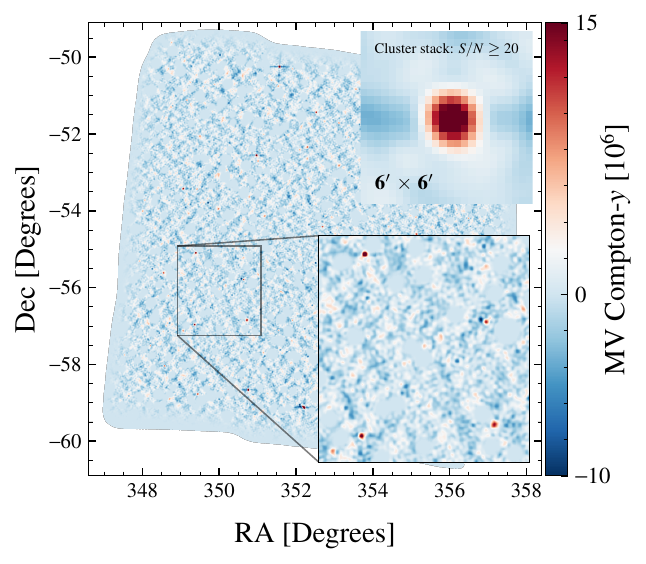}
\caption{MV Compton-$y$ map created using the full dataset. 
The increments (red dots) correspond to the cluster locations and the negative bowls around them correspond to the scan-direction filtering of the SPT maps, which we correct using TF during the power spectrum estimation stage. 
%The isolated decrements (blue dots) correspond to the location of radio Poisson sources that are below our detection threshold and hence unmasked. 
The inset panel in the top right shows the $\clusterstacksize$ stack of clusters with $\snr \ge 20$ from \citet{kornoelje25}. 
%This map is affected by the noise bias and is intended solely for illustrative purposes.
}
\label{fig_ymap_data}
\end{figure}

\begin{figure*}
\centering
\includegraphics[width=0.9\textwidth, keepaspectratio]{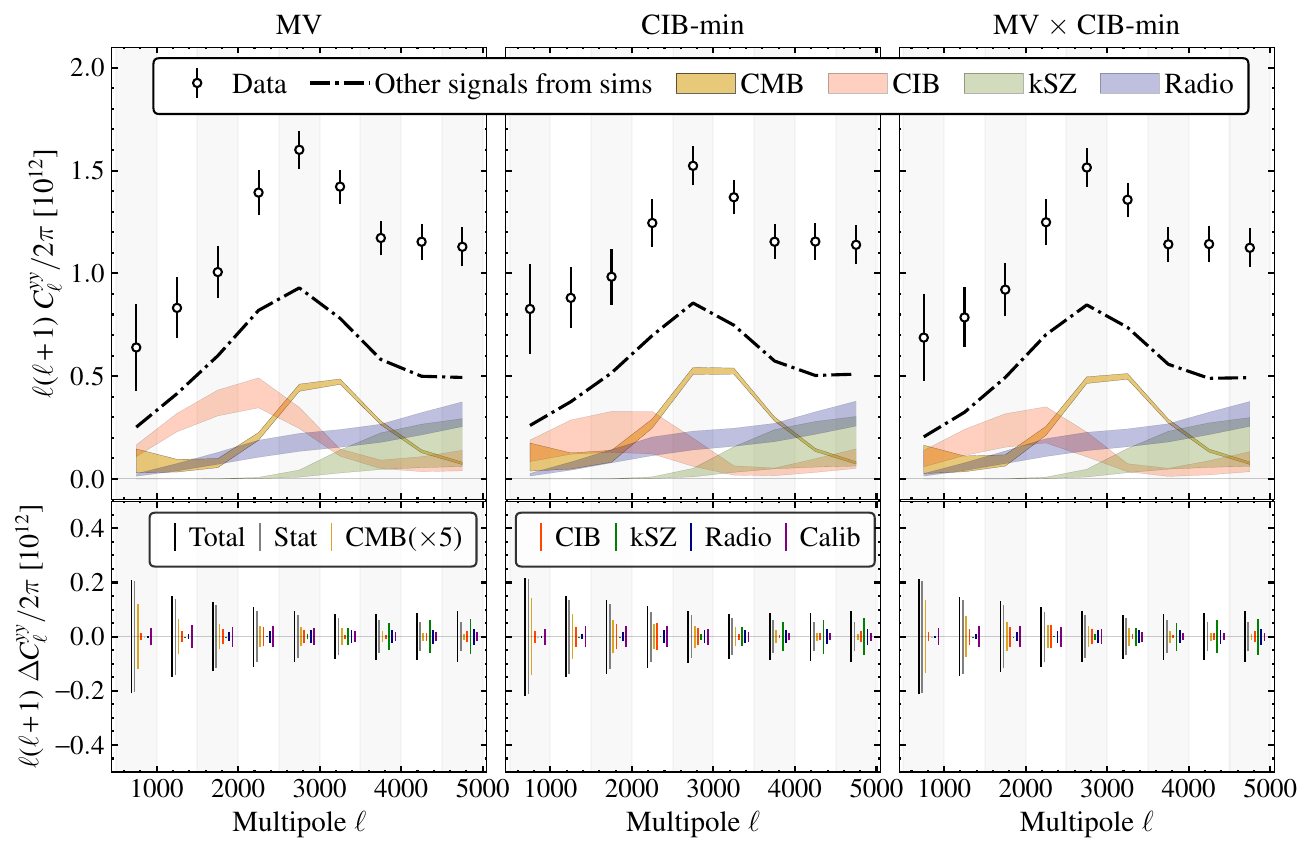}
\caption{The total measured power spectra from data are shown as open circles in top panels for MV (left), \cibfree{} (middle), and the cross-spectrum between the two (right). 
The dash-dotted line corresponds to the expectation spectra of the undesired components with the individual contributions shown in different colors: CMB in yellow, CIB in red, kSZ in green and radio in blue. 
The dash‑dotted curve does not have the tSZ contribution, so it is not expected to align with the data points.
The bands correspond to the systematic scatter in each component (see text in \S\ref{sec_sys_errors} for details). 
The error budgets (black) are presented in the bottom panels and have been decomposed into statistical (gray) and systematic errors from the undesired signals and calibration uncertainties (all other colors).
}
\label{ymap_spectra_data_simulations_sys_checks_plus_errors}
\end{figure*}

\begin{figure*}
\centering
\includegraphics[width=0.9\textwidth, keepaspectratio]{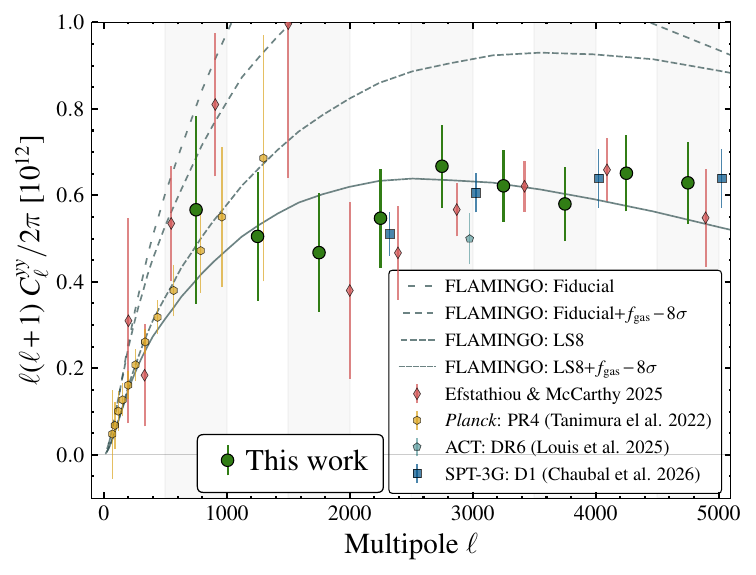}
\caption{The tSZ power spectra measured at $\tszpowerspectrumsnr$ using the \cibfree{} maps is shown in green. 
The error bars include contributions from both statistical and systematic errors, and have significant off-diagonal correlations due to the non-Gaussian nature of the signals. 
We also show the results from \planck{} (PR4) as yellow hexagons; ACT (DR6) as a teal octagon; \sptthreeg{} (D1) as blue squares; and a joint analysis of ACT, \planck, and SPT as red diamonds.
While the results in this work are slightly higher than ACT at $\ell = 3000$ by $1\sigma$, we find excellent agreement with \planck{} in the overlapping scales and also with the results from \sptthreeg{} and \citet{efstathiou25} over a wide range of scales. 
%For reference, we also show the power spectra from versions of the \flamingo{} simulations as gray curves: the dotted curve is for the low $\seight$ (LS8) while the dashed curve corresponds to the model with low $\seight$ and with high ($f_{\rm gas} - 8\sigma$) AGN feedback.
%\refresponse{
For reference, we also show the power spectra from versions of the \flamingo{} simulations as gray curves with different spacings. The curve with the widest spacing corresponds to the simulation with the fiducial cosmology and fiducial AGN feedback. 
The next curve (with slightly smaller spacing) shows the same cosmology but with high AGN feedback ($f_{\rm gas} - 8\sigma$). 
This is followed by the low $\seight$ (LS8) cosmology with fiducial feedback, and finally the model combining low $\seight$ with high AGN feedback (LS8+$f_{\rm gas} - 8\sigma$).} %}
\label{fig_ymap_spectra_data_plus_literature}
\end{figure*}

Following the discussion in \S\ref{sec_map_bundles_cross_spectra}, we create a number of Compton-$y$ maps from data and simulations as follows:
\begin{itemize}
\item{{\bf Data:} Two\footnote{As mentioned earlier in \S\ref{sec_ilc}, we do not consider CMB-free and \radiofree{} maps for the power spectra estimation due to high levels of noise and CIB compared to the MV and \cibfree{} maps.} Compton-$y$ maps ---MV and \cibfree--- for all the hundred data map bundles. In total, we have 100 (bundles) $\times$ 2 (LC) = 200 maps.

In Fig.~\ref{fig_ymap_data}, we also show the MV Compton-$y$ map reconstructed using the full dataset. 
The inset in top right shows the $\clusterstacksize$ stack at the locations of clusters detected with $\snr$ $\ge$ 20 from \citet{kornoelje25}. 
Note that we only use this full-coadd map for plotting; all power spectrum analyses are performed using cross-spectra of partial-depth maps.
%Since this map, created from the full dataset, is affected by noise bias, we present it solely for illustrative purposes and do not use it for any quantitative analysis.
}
\item{{\bf Simulations:}}
	\begin{itemize}
		\item{Set-A: For the systematic checks and to compute the expectation spectra from the undesired components, we use \howmanyskypatches{} \agora{} simulations. For each simulation, to remove the noise bias in the power spectra, we create two sets of Compton-$y$ maps with independent noise realizations. In total we have \howmanyskypatches{} (realizations) $\times$ 2 (LC) $\times$ 2 (bundles) = 100 maps. We find \howmanyskypatches{} realizations to be more than sufficient to reduce the sample variance since the templates converge after $\sim10$ realizations.}
		\item{Set-B: To estimate the bandpower covariance matrix, we obtain the MV and \cibfree{} maps from 100 \agora{} simulations that include experimental noise. Here, we have 100 (realizations) $\times$ 2 (LC) = 200 maps. This covariance matrix ---distinct from the one used to obtain the LC weights in \S\ref{sec_ilc}--- is used to capture both the measurement uncertainties and the correlations between adjacent $\ell$ bins of the tSZ power spectrum.}
	\end{itemize}
\end{itemize}
We compute the cross power spectra from the above data maps and Set-A simulations according to Eq.(\ref{eq_cross_spectra}). 
To get the bandpower covariance matrix, we use the auto power spectra of the Set-B simulations. 
As described in \S\ref{sec_map_bundles_cross_spectra}, before computing the spectra, we mask the location of sources (which have been inpainted in the individual frequency maps) in the Compton-$y$ maps. 

In the top panels of Fig.~\ref{ymap_spectra_data_simulations_sys_checks_plus_errors}, we present the auto-power spectra for the MV (left) and \cibfree{} (middle) maps, and the cross-spectra between the two maps (right).
The open circles show the power spectra measured from data bundles. 
The black dash-dotted curves represent the expectation spectra for the undesired components ---CMB, CIB, radio and kSZ--- and have been averaged over \howmanyskypatches{} simulations. 
Note that these expectation spectra do not contain the tSZ signal and hence are not expected to align with the data shown as open circles. 
To be more specific, we compute the expectation spectra as the difference in power spectra obtained from (a) \agora{} simulations containing CMB, \texttt{Uncorr-} CIB, \texttt{Uncorr-} radio, kSZ, tSZ and (b) tSZ-only \agora{} simulations. 
As mentioned above, we break the correlations between tSZ and CIB/radio in simulations when calculating the expectation spectra and we discuss more about this in \S\ref{sec_tsz_cib}. 
We adopt this procedure rather than computing the expectation spectra from \agora{} simulations that only contain CMB, CIB, radio and kSZ to ensure that the statistical properties of the simulations match the data as closely as possible. 
To get the final tSZ-only spectra, we subtract the black dash-dotted curve from the data points; this step is discussed in more detail in the subsequent sections.

In the bottom panels of Fig.~\ref{ymap_spectra_data_simulations_sys_checks_plus_errors}, we show the total bandpower errors (black) that include both the statistical (gray) errors and the systematic errors (other colors), which are estimated as described in Appendix~\ref{sec_sys_errors}. %below. 
For ease of comparison, we have centered the errors on zero.

\subsection{Bandpower errors and covariance matrix}
\label{sec_bandpower_errors}
We obtain the bandpower covariance matrix ${\bf C}_{\rm Stat}$ using the \howmanysimulations{} Set-B \agora{} simulations (see \S\ref{sec_ymaps_power_spectrum}) for MV, \cibfree, and the cross-power spectrum between the two. 
Since the tSZ power spectrum is dominated by massive clusters that remain unmasked in this analysis, we expect the errors to be non-Gaussian leading to off-diagonal correlations in the covariance matrix \citep{seljak00}. 
The non-Gaussian \agora{} simulations should properly capture these correlations and also the sample variance from the tSZ signal. 
%\refresponse{
However, we note that \agora{} is a single simulation; consequently, it does not capture uncertainties arising from the underlying cosmological model or from baryonic physics, nor does it account for the possible dependence of baryonic processes on cosmology. %}
%We also compute the systematic covariance matrix ${\bf C}_{\rm Sys}$ using the results from the previous section. 
We also compute the systematic covariance matrix, ${\bf C}_{\rm Sys}$, using the methods and results from Appendix~\ref{sec_sys_errors}. 
Specifically, we compute ${\bf C}_{\rm Sys}$ for calibration errors, CMB, CIB, kSZ and radio using the scatter shown as bands in Fig.~\ref{ymap_spectra_data_simulations_sys_checks_plus_errors}. 
The total covariance matrix is then 
\begin{eqnarray}
\label{eq_total_cov} 
{\bf C}_{\rm Total} = {\bf C}_{\rm Stat} + {\bf C}_{\rm Calib-Sys}  + {\bf C}_{\rm CMB-Sys} + \\\notag
{\bf C}_{\rm CIB-Sys} + {\bf C}_{\rm kSZ-Sys} + {\bf C}_{\rm Rad-Sys}
\end{eqnarray}
and we show the square root of the diagonal of the total (black) and the individual components (Stat - gray; CMB - \cmbresidualcolour; CIB - \cibresidualcolour; kSZ - \kszresidualcolour; Radio - \radioresidualcolour; Calib - purple) in the lower panels of Fig.~\ref{ymap_spectra_data_simulations_sys_checks_plus_errors} for MV (left), \cibfree{} (middle), and MV $\times$ \cibfree{} (right). 
%For clarity, we enhance the systematic error due to CMB by $\times5$. 
For illustrative purposes, we enhance the systematic error due to CMB by $\times5$ to demonstrate the relative size of error as a function of scale.
We note that the large scales are dominated by the statistical errors (mainly the sample variance) and the systematics errors become comparable to the statistical errors on small scales.  
The systematic errors are highly correlated between adjacent bins and this is taken into account in the covariance matrix.

\subsection{tSZ power spectrum estimate}
\label{sec_tsz_only_power_spectrum}
In this section, we report the final tSZ power spectrum estimates from this work and compare the results with the measurements reported by different surveys. 

The green circles in Fig.~\ref{fig_ymap_spectra_data_plus_literature} show the tSZ power spectrum measurements obtained here, derived from the cross‑spectra of the 
\howmanybundles\ \cibfree{} Compton‑$y$ map bundles. 
As mentioned in \S\ref{sec_map_bundles_cross_spectra}, we do not compute the odd-odd or even-even cross-spectra since they have the same \herschel-SPIRE splits and will be affected by the noise bias. 
The measurement significance is $\tszpowerspectrumsnrforcibfree$ in the multipole range $\ell \in [\lminforps, \lmaxforps]$. 
The $\snr$ drops by $20-27\%$ when reducing the small scale limit to $\ellmax = 4000, 3500$, and $3000$. 

We adopt \cibfree{} as our fiducial estimate because the residual CIB bias arising from the tSZ–CIB cross‑correlation has a minor impact on the large scales in both MV and \mvcrosscibfree. 
In the next subsection (\S\ref{sec_tsz_pow_spec_ilc_and_tsz_cib}), we provide a detailed comparison of the tSZ estimates obtained from all three LC combinations (MV, \cibfree, and \mvcrosscibfree), as well as discuss the scale-dependent reconstruction of \tszcrosscib. 
%In Appendix~\ref{sec_tsz_pow_spec_ilc_and_tsz_cib} and Fig.~\ref{fig_ymap_spectra_data_plus_simulations}, we provide a detailed comparison of the tSZ estimates obtained from all three LC combinations (MV, \cibfree, and \mvcrosscibfree), as well as the scale-dependent reconstruction of \tszcrosscib{} presented in Appendix~\ref{sec_tsz_cib}. To summarize, we find excellent agreement between all the three estimates at $\ell \gtrsim 2500$. There are noticeable differences on large scales, and we use these differences to reconstruct $\rhotszcrosscib$ as a function of multipole. We find a $\snrofrhotszcibformvandcibfree\sigma$ preference for non-zero correlation, with a preferred value of the correlation coefficient of $\sim 0.2$ on large scales, decreasing rapidly towards zero at $\ell \gtrsim 2500$. The measurement significance of $\rhotszcrosscib$ is between $2$-$3\sigma$ depending on the LC combinations used for the difference test. More details are provided in Appendix~\ref{sec_tsz_pow_spec_ilc_and_tsz_cib} and in Figs.~\ref{fig_rho_tsz_cib} and \ref{fig_rho_tsz_cib_corner}. 

In Fig.~\ref{fig_ymap_spectra_data_plus_literature}, we also show measurements reported by different groups in the literature: \planck{} (PR4) as yellow hexagons \citep{tanimura22}; ACT (DR6) as a teal octagon \citep{louis25}; %SPT as an orange octagon \citepalias{reichardt21}; 
and the measurement from a joint analysis of \planck, ACT, and SPT data as red diamonds \citep{efstathiou25}. 
We also show the recent results from \sptthreeg{} (D1) \citep{chaubal26} as blue squares. 
The contribution from other components are either removed or taken into account using multi-component fits in all these works.
%\refresponse{
The error bars in the current work are much larger than those reported in \citet{chaubal26}. The primary reason is sky coverage: our analysis uses a sky area roughly 15 times smaller than that of \citet{chaubal26}. However, additional factors also contribute to the differences in the errors between the two works. In the present work, we include the non‑Gaussian contribution to the tSZ covariance, which is non‑negligible, whereas \citet{chaubal26} excluded this term. Moreover, \citet{chaubal26} incorporates small-scale information up to $\ell = 11,000$, which provides stronger constraints on the Poisson foregrounds --namely the CIB and radio galaxies. On the other hand, our use of \herschel-SPIRE data offers improved control of the CIB component relative to \citet{chaubal26}. %}
%However, the approach to handling systematic errors arising from assumptions about other components is less clear. Our results are higher than the ACT and SPT measurements at $\ell = 3000$ by $1-2\sigma$. 
Our results are higher than the ACT measurements at $\ell = 3000$ by $\sim1\sigma$. 
However, we find excellent agreement with \planck{} data in the overlapping scales $\ell \in [500, 1100]$ and also with \sptthreeg{} (D1) at $\ell \ge 2000$. 
Our results are also in good agreement with the results of \citet{efstathiou25}, %except for a $1\sigma$ deviation at $\ell \sim 2200$, 
over a wide range of scales, particularly at $\ell \ge 2000$. 

For reference, we also plot the tSZ power spectrum from \flamingo{} simulations \citep[Fig. 6 of][]{mccarthy25} as gray curves. 
We emphasize that these curves are not fits to our measurements but are shown solely for illustrative purposes. 
%\refresponse{
The fiducial \flamingo{} cosmology corresponds to the parameters obtained by combining the $3 \times 2$ measurements from the Dark Energy Survey (DES) with CMB data from \planck. 
This also includes information from additional external datasets --baryonic acoustic oscillations, redshift-space distortions, and Type-Ia supernovae. 
They also provide a second simulation run with a lower value of $\seight = 0.766$, derived solely from DES lensing measurements combined with \planck{} CMB data; this model is referred to as LS8. 
The curve with the widest spacing corresponds to the simulation using the fiducial cosmology and the fiducial level of Active Galactic Nucleus (AGN) feedback. This is followed by the curve with the same cosmology but with a high-feedback model ($f_{\rm gas} - 8\sigma$), representing feedback increased by eight times the calibration uncertainty in the simulation. 
As noted by \citet{mccarthy25}, both of these models substantially over predict the tSZ power spectrum on large angular scales relative to the \planck{} measurements (yellow hexagons). However, several of the large-scale data points from \citet{efstathiou25} are consistent with these predictions, as is the first multipole bin from this work. 
On the other hand, as demonstrated by \citet{mccarthy25}, the LS8 cosmology provides a better fit to the large-scale \planck{} measurements than the fiducial value of $\seight = 0.833$. However, this model still over predicts the tSZ power spectrum on small scales relative to the data. The LS8+$f_{\rm gas}-8\sigma$ model, shown as the gray curve with the tightest spacing, offers a better match to our measurements as well as to other recent results on small scales. %}
%As inferred by \citet{mccarthy25}, the LS8 cosmology model (dotted gray curve) with a lower value of $\seight = 0.766$ but with the fiducial Active Galactic Nucleus (AGN) feedback model, derived based on joint analysis of galaxy clustering and weak lensing measurements, is a better fit to the large-scale measurements of the tSZ power spectrum compared to the fiducial value of $\seight = 0.833$ from \planck{} CMB measurements. However, this model over predicts the tSZ power spectrum on small-scales compared to the measurements as can be seen in the figure and also shown by \citet{mccarthy25, efstathiou25}. On the other hand, a tuned LS8 model with a high level (a model with feedback higher than the fiducial level by eight times the calibration uncertainty in the simulation) of feedback, LS8+$f_{\rm gas}-8\sigma$, shown as the gray dashed curve, matches our results better. 
Note that all these models converge on large-scales, matching the \planck{} measurements, since those scales are dominated by massive, low‑redshift clusters, which are less affected by feedback processes \citep{mccarthy25}.

\subsection{tSZ power spectrum from different LC techniques and \tszcrosscib{} reconstruction}
\label{sec_tsz_pow_spec_ilc_and_tsz_cib}
\begin{figure*}
\centering
\includegraphics[width=\textwidth, keepaspectratio]{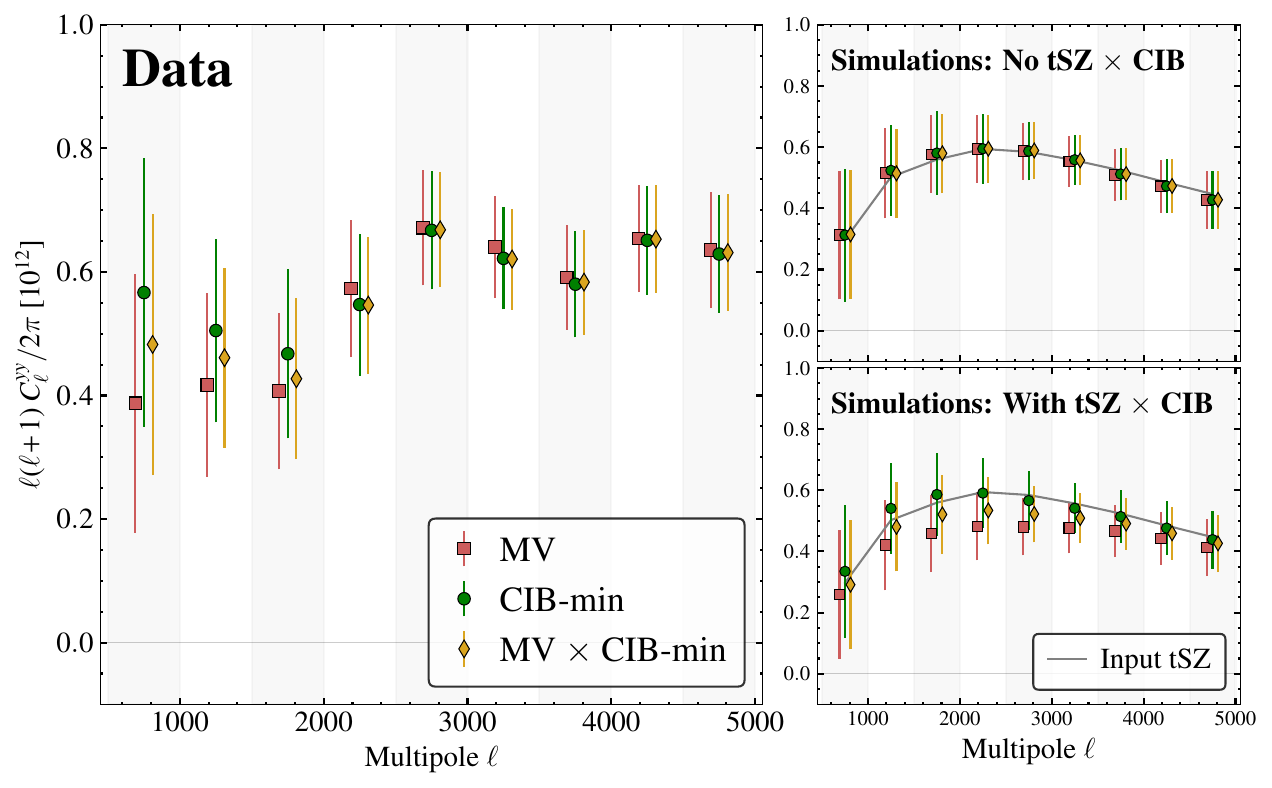}
\caption{Reconstructed tSZ power spectra measured from data (left) and simulations (right): MV (red squares), \cibfree{} (green circles), and their cross-spectrum (yellow diamonds).
{\it Left panel:} In data, the tSZ power spectrum is measured at $\sim \tszpowerspectrumsnr$ in all the three cases. 
At $\ell \gtrsim 2500$, we find excellent agreement between all three spectra. 
The differences at lower multipoles are due to the different levels of \tszcrosscib, which results in a partial cancellation of the recovered tSZ signal. 
{\it Right panels:} The input tSZ power spectrum in simulations is show as the gray curve. 
The top right panel corresponds to \agora{} simulations with all the components but after explicitly removing the \tszcrosscib{} using (CMB + kSZ + tSZ + \texttt{Uncorr-}CIB + \texttt{Uncorr-}radio). 
In the absence of \tszcrosscib, all of the estimators return unbiased results. 
The bottom right panel contains \tszcrosscib, and in this case, we find the MV estimate to be biased low at large scales. 
The \cibfree{} remains unbiased and \mvcrosscibfree{} lies in between the two. 
The simulations results are obtained after averaging over \howmanyskypatches{} realizations. 
This trend is similar between data and simulations, although unlike in the empirical data, the simulations show that MV remains biased even at small scales.
This is due to the differences in \tszcrosscib{} between data and simulations. 
The bottom right panel is provided solely for illustration to build intuition about how the \tszcrosscib{} can bias the tSZ power spectrum, and we emphasize that we do not rely on \tszcrosscib{} in simulations for any part of our analysis.
}
\label{fig_ymap_spectra_data_plus_simulations}
\end{figure*}

In \S\ref{sec_tsz_only_power_spectrum}, we presented the tSZ power spectrum results derived from the \cibfree{} technique and compared it with other works from the literature. 
In the left panel of Fig.~\ref{fig_ymap_spectra_data_plus_simulations}, we show and compare the results from different LC combinations: MV (red squares), \cibfree{} (green circles), and the cross-spectrum of the two (yellow diamonds). 
These are obtained by subtracting the expectation spectra of the undesired signals (shown as the black dash dotted curve in Fig.~\ref{ymap_spectra_data_simulations_sys_checks_plus_errors}) from the total power spectra (open circles in Fig.~\ref{ymap_spectra_data_simulations_sys_checks_plus_errors}). 
The systematic errors in the expectation spectra, described above, are taken into account in the errors.
As mentioned in the main text, the adjacent bins are correlated due to the non-Gaussian nature of the signals.
We measure the tSZ power spectra in the multipole range $\ell \in [\lminforps, \lmaxforps]$ at $\tszpowerspectrumsnrformv$ with MV, $\tszpowerspectrumsnrforcibfree$ with \cibfree, and $\tszpowerspectrumsnrformvcrosscibfree$ using the MV $\times$ \cibfree{} cross spectrum.
There is an excellent agreement between all three spectra at scales $\ell \gtrsim 2500$. 
At large scales, \cibfree{} data points in green are slightly higher in the range $\ell \in [1500, 2500]$. 
This is due to the partial cancellation of the tSZ signal and the residual CIB caused by non-zero \tszcrosscib, which impacts the MV estimate more. 
In this work, we reconstruct the scale‑dependent \tszcrosscib, since, beyond the challenges of accurately modeling the CIB, the \tszcrosscib{} correlation itself remains a major obstacle to extracting robust information from both the kSZ and tSZ signals (\citetalias{reichardt21, raghunathan23}; \citealt{tristram25}).

To build intuition about the impact of \tszcrosscib{} on the reconstructed tSZ power spectra, we present the simulation results, averaged over \howmanyskypatches{} simulations, in the right panels of Fig.~\ref{fig_ymap_spectra_data_plus_simulations}. 
In the top right panel, we show results that include all the components from \agora{} simulations but after explicitly removing the correlation between tSZ and CIB, (i.e:) CMB + kSZ + tSZ + \texttt{Uncorr-}CIB + \texttt{Uncorr-}radio. 
In this case, the recovered tSZ power spectrum estimates for all three cases match the input tSZ, shown as the solid gray curve. 
This further demonstrates that our pipeline yields unbiased results. 
In contrast, when \tszcrosscib{} is present (bottom right panel), the residual CIB contributions partially cancel the tSZ signal, resulting in a biased estimate. 
This bias is most pronounced for MV and somewhat reduced for \mvcrosscibfree, while the \cibfree{} method remains unbiased. 
 The trend in tSZ amplitude among the different LC estimates is similar in the data (left) and the simulations (bottom right). 
However, we note that in the simulations, the MV estimate differs from the others over all relevant scales, while in the data the three LC estimates agree at $\ell \gtrsim 2500$. 
This difference between the data and simulations arises from differences in the \tszcrosscib{} contribution. 
Since the scale dependence of the \tszcrosscib{} has not been detected at high significance to date, this correlation is not properly calibrated in the simulations. 
Consequently, we do not use \tszcrosscib{} in the simulations in our analysis.

\subsubsection{Reconstructing the cross-correlation between tSZ and CIB}
\label{sec_tsz_cib}

\begin{figure}
\centering
\includegraphics[width=0.48\textwidth, keepaspectratio]{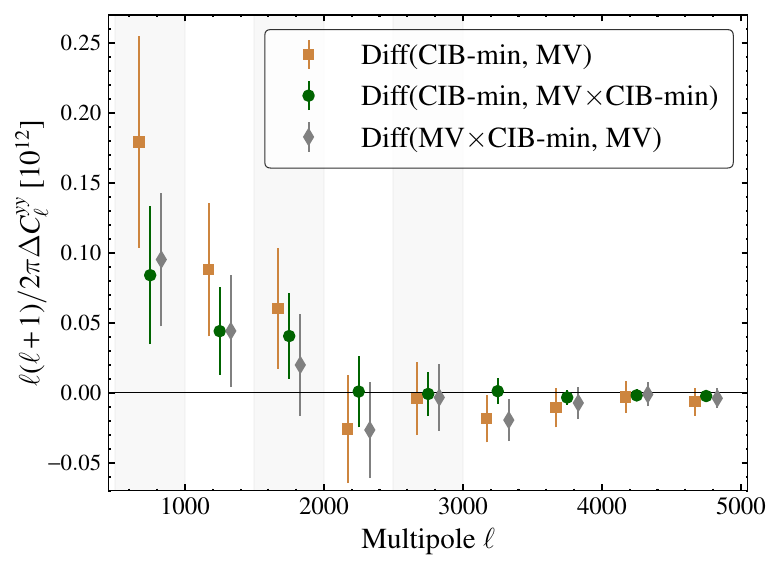}
\caption{The difference spectrum $\Delta C_{\ell_{XY}}^{yy}$ computed from different LC combinations are shown. The difference spectrum should be consistent with zero in the absence of any bias and we attribute the differences to the residual CIB bias arising due to \tszcrosscib. 
The error bars are computed using the square root of the diagonal of the covariance matrix of the difference spectrum in Eq.(\ref{eq_cov_difference_spectrum}). 
As is the case for $\clyy$, the adjacent $\ell$ bins are highly correlated. 
}
\label{fig_difference_spectrum}
%\vskip -10pt
\end{figure}

\begin{figure}
\centering
\includegraphics[width=0.48\textwidth, keepaspectratio]{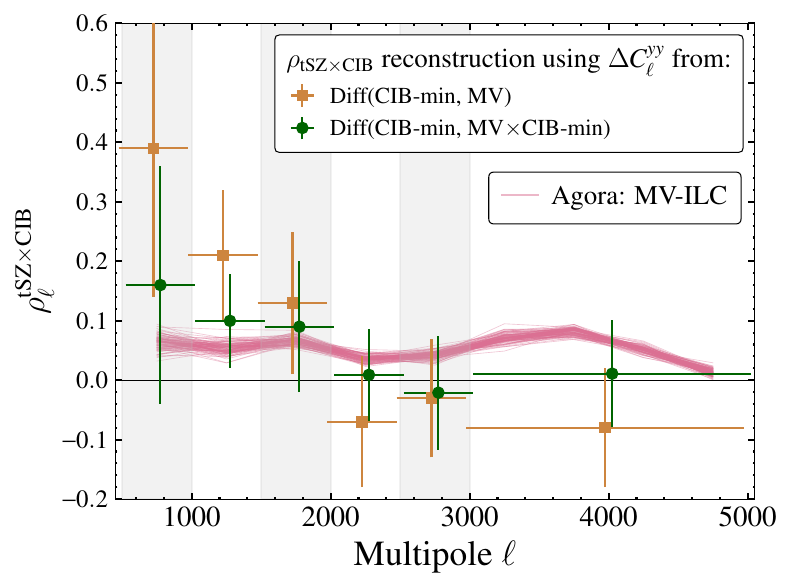}
%\includegraphics[trim=0.in 0.in 0in 0.0in,clip=true, width=0.5\textwidth, keepaspectratio]{figs/rho_tsz_cib.pdf}
% \hspace{-0.9cm}
% \includegraphics[trim=0.in 0.in 0in 0.0in,clip=true, width=0.55\textwidth, keepaspectratio]{figs/rho_tsz_cib_corner_6bins.pdf}
\caption{Reconstructed scale-dependent $\rhotszcrosscib$ using the difference spectrum from two LC combinations: (\cibfree, MV) as yellow squares and (\cibfree, \mvcrosscibfree) as green circles. Combining all the bins, which are highly correlated, we find $\snrofrhotszcibformvandcibfree\sigma$ and $\snrofrhotszcibforcibfreeandmvcrosscibfree\sigma$ for a non-zero $\rhotszcrosscib$ from the two cases. 
For illustrative purposes, we additionally present the $\rhotszcrosscib$ inferred from 
%\refresponse{\howmanysimulations{} \agora{} MV-LC simulations}, 
\howmanysimulations{} \agora{} MV-LC simulations, 
shown in pink, although it is not used in any part of our analysis. 
The full corner plot is presented in Fig.~\ref{fig_rho_tsz_cib_corner}.}
\label{fig_rho_tsz_cib}
%\vskip -10pt
\end{figure}

\begin{figure*}
\centering
\includegraphics[width=0.8\textwidth, keepaspectratio]{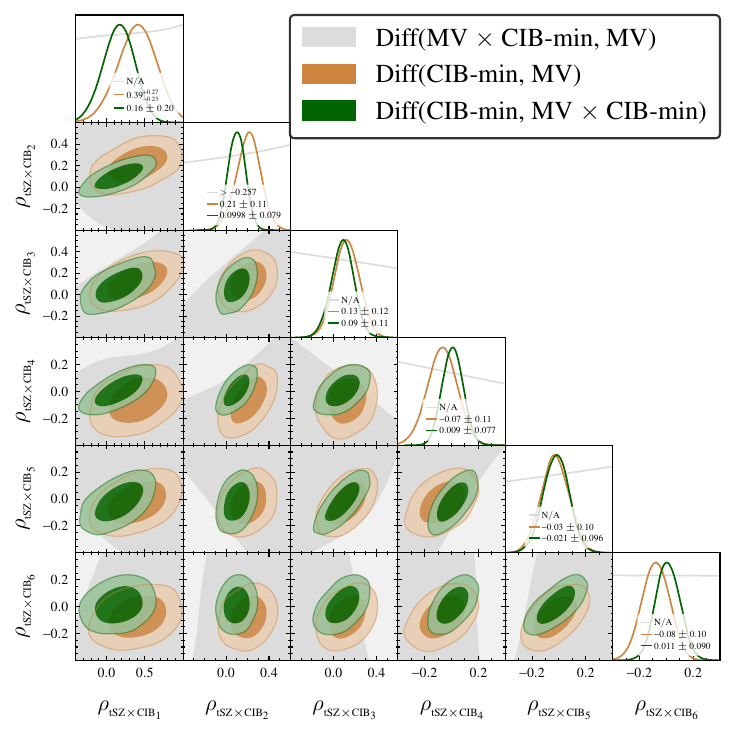}
\caption{Full corner plot showing the correlations between $\rhotszcrosscib$ in the six bins shown in Fig.~\ref{fig_rho_tsz_cib}. 
The colors follow the same scheme as Fig.~\ref{fig_difference_spectrum} and Fig.~\ref{fig_rho_tsz_cib}. 
For completeness, we also include in gray the results from the difference test of \mbox{(\mvcrosscibfree, MV);} however, this test provides no additional constraining power.
} 
\label{fig_rho_tsz_cib_corner}
\end{figure*}
Based on the data and simulation results shown in Fig.~\ref{fig_ymap_spectra_data_plus_simulations}, we attribute the differences among the reconstructed tSZ power spectra from the three LC estimates to \tszcrosscib{}, and we use these differences to reconstruct the \tszcrosscib{} cross‑correlation.
To this end, we compute the difference spectrum between two tSZ power spectrum estimates \mbox{$X$ and $Y$}
\begin{equation}
\Delta C_{\ell_{XY}}^{yy} = C_{\ell_{\rm X}}^{yy} - C_{\ell_{\rm Y}}^{yy}, 
\end{equation}
where $X, Y$ correspond to the power spectrum of different LC estimates. 
Specifically, they correspond to auto-spectrum $\alpha \alpha$ or the cross-spectrum $\alpha \beta$ of the LC estimates with $\alpha, \beta \in$ [MV, \cibfree]. 
The covariance of this difference spectrum is given as 
\begin{equation}
{\bf C}_{XY} = {\bf C}_{\alpha\alpha} + {\bf C}_{\beta\beta} - 2{\bf C}_{\alpha \beta}.
\label{eq_cov_difference_spectrum}
\end{equation}

We show $\Delta C_{\ell_{XY}}^{yy}$ computed for different combinations in Fig.~\ref{fig_difference_spectrum}: green circles are for $(X, Y)=$ (\cibfree, \mvcrosscibfree); yellow squares are for $(X, Y)=$ (\cibfree, MV); and gray diamonds are for $(X, Y)=$ (\mvcrosscibfree, MV). 
In the absence of any bias, $\Delta C_{\ell_{XY}}^{yy}$ is expected to be consistent with a null spectrum within the errors. 
We attribute the bias to the cancellation of the tSZ signal by the residual CIB present in the Compton-$y$ maps arising due to \tszcrosscib{} and use this information to study $\rhotszcrosscib$. 

We reconstruct $\rhotszcrosscib$ in multiple $\ell$ bins to understand how the cross-correlation changes with scales. 
We adopt six bins in our fiducial setup: (a) five bins with $\Delta \ell = 500$ for $\ell \in [500, 3000]$ and (b) a single large bin for $\ell \in [3000, 5000]$.
The binning choice is also highlighted as light gray shaded regions in Fig.~\ref{fig_difference_spectrum}. 
There are negligible changes to the results when we use the same binning used for $\clyy$ with $\Delta \ell = 500$ for $\ell \in [500, 5000]$. 
This is because the difference spectra $\Delta C_{\ell_{XY}}^{yy}$ are consistent with zero at $\ell \gtrsim 2500$. 
The parameter estimation is carried out using a Markov Chain Monte Carlo (MCMC) framework, utilizing the Code for BAYesian Analysis (\texttt{cobaya}\footnote{\url{https://cobaya.readthedocs.io/en/latest/}}, \citealt{torrado21}) to sample the posterior distributions with the Metropolis-Hastings sampler. 
The chains are analyzed using the \texttt{GetDist}\footnote{\url{https://getdist.readthedocs.io/en/latest/index.html}} package and we set the Gelman-Rubin statistic \mbox{$R-1 = 0.01$} implemented in \texttt{cobaya} based on \citet{lewis13} to attain the chain convergence. 

We perform the fitting in power spectrum space (using the beam and transfer function deconvolved results). 
%\srini{Working on setting up the map-based approach since the Cl approach cannot distinguish between correlated vs uncorrelated CIB. It is extremely computationally intensive, though. It needs  tSZ/CIB map simulations for all six bands with varying levels of \tszcrosscib{} in six bins. The simulations need to repeated for every step in the MCMC process and so unless I build an emulator, this is almost impossible. Given the results are only $\snrofrhotszcibformvandcibfree\sigma$, I may give up on this.}
We compare the data vectors $\Delta C_{\ell_{XY}}^{yy}$ obtained from the differences of the LC combinations to models computed for the corresponding LC combinations as follows. 
Given the power spectra of tSZ $C_{\ell_{\nu_1}}^{\rm tSZ}$ and CIB $C_{\ell_{\nu_2}}^{\rm CIB}$ in two frequency bands $\nu_{1}$ and $\nu_{2}$, we define the \tszcrosscib{} cross-spectrum $C_{\ell_{\nu_1 \nu_2}}^{\rm tSZ x CIB}$ using a scale-dependent correlation coefficient $\rhotszcrosscib$ as
\begin{eqnarray}
C_{\ell_{\nu_1 \nu_2}}^{\rm tSZ \times CIB} = & -\rho_{\ell}^{\rm tSZ \times CIB} \sqrt{C_{\ell_{\nu_1}}^{\rm CIB} C_{\ell_{\nu_2}}^{\rm tSZ}}
\label{eq_tsz_cib}
\end{eqnarray} 
where $\nu_{1}, \nu_{2}$ correspond to one of the six frequency bands from the \sptthreeg, SPTpol, and \herschel-SPIRE surveys. 

In Eq.(\ref{eq_tsz_cib}), the CIB power spectra for different bands come from \agora{} simulations. 
For the tSZ power spectra, we follow two approaches. 
In the first approach, we use the Compton-$y$ power spectrum derived from \cibfree{} LC estimates (green points in Fig.~\ref{fig_ymap_spectra_data_plus_literature}) and multiply the spectrum by the tSZ frequency dependence to get to the $\Delta T_{\rm CMB}$ units in $\mu K$. 
In the second approach, we use the tSZ power spectra from \agora{} simulations. 
The first approach is appealing because it is independent of amplitude and shape differences between the true tSZ power spectrum and the simulations. 
However, this approach implicitly assumes that the bias due to \tszcrosscib{} is zero in the \cibfree{} estimate. 
The second approach, on the other hand, is immune to the bias from \tszcrosscib{}, but it can be affected by mismatches between the true underlying tSZ power spectrum and the simulations. 
Thus, the two approaches are complementary, and our results change only minimally between the two cases. 
Note that in both approaches, we do not use the \tszcrosscib{} in simulations and inject our own scale-dependent correlation based on Eq.(\ref{eq_tsz_cib}). 

We apply the LC weights for two different Compton-$y$ estimates $w_{\ell_{\alpha}}$ and $w_{\ell_{\beta}}$ to $C_{\ell_{\nu_1 \nu_2}}^{\rm tSZ x CIB}$ and obtain the final LCed estimate $C_{\ell_{\rm \alpha \beta}}^{\rm tSZ \times CIB}$ as given below:
\begin{eqnarray}
C_{\ell_{\rm \alpha \beta}}^{\rm tSZ \times CIB} = & w_{\ell_{\alpha}} {\bf C}_{\ell}^{{\rm tSZ \times CIB}}\ w_{\ell_{\beta}}^{\dagger}. 
\label{eq_tsz_cib_ilc}
\end{eqnarray}
Finally, we compute the model vector as 
\begin{eqnarray}
    \Delta C_{\ell_{XY, {\rm model}}}^{yy} = & 2 (C_{\ell_{X}}^{\rm tSZ \times CIB} - C_{\ell_{Y}}^{\rm tSZ \times CIB}). 
\label{eq_tsz_cib_model_vector}
\end{eqnarray}
To reduce scatter from a single simulation, we apply the above operations to \howmanyskypatches{} simulations and adopt the resulting average as our final model vector. 

In our baseline case, we have six free parameters corresponding to $\rho_{\ell_{i}}^{\rm tSZ \times CIB}$ in the six $\ell_{i}$ bins described above. 
We assume uniform priors $\mathcal{U}(-1, 1)$ for all bins. 
We present the results in Fig.~\ref{fig_rho_tsz_cib} which shows the best-fit results for $(X, Y)=$ (MV, \cibfree) as yellow squares and $(X, Y)=$ (\cibfree, \mvcrosscibfree) as green circles. 
Since the \tszcrosscib{} partially cancels in $(X, Y)=$ (\mvcrosscibfree, MV), that test is not as constraining as the other two tests, as shown in Fig.~\ref{fig_rho_tsz_cib_corner}.
The cross-correlation coefficient is non-zero on large scales ($\ell < 2500$), but it decreases and approaches zero at smaller scales.
Combining all six bins, we obtain $\snrofrhotszcibformvandcibfree \sigma$ and $\snrofrhotszcibforcibfreeandmvcrosscibfree \sigma$ evidence for non-zero $\rhotszcrosscib$ for the two cases presented above,
%: $(X, Y)=$ (MV, \cibfree) in yellow and  $(X, Y)=$ (\cibfree, \mvcrosscibfree) in green, 
and they both agree well with each other.
%For reference, we also show the $\rhotszcrosscib$ computed from \agora{} simulations as the pink curve. 
%\refresponse{
For reference, we also show the $\rhotszcrosscib$ computed from the \howmanysimulations{} \agora{} MV-LC simulations in pink. Unlike the case for the data, the correlation coefficient for \agora{} is non-zero at small scales and is broadly consistent with the expectations from the bottom right panel of Fig.~\ref{fig_ymap_spectra_data_plus_simulations}, which shows that the MV reconstruction (red squares) is biased across all scales. %}
We do not use \agora{} $\rhotszcrosscib$ in any part of our analysis. 
These results correspond to approach (A) where the tSZ power spectra used for modeling $C_{\ell_{\nu_1 \nu_2}}^{\rm tSZ \times CIB}$ are obtained from \cibfree{} results. 
We find negligible impact on the results when we switched to the tSZ power spectrum from \agora{} simulations (Approach B). 
We also checked the impact of beam uncertainties (\S\ref{sec_beam_uncertainities}) and also find them to be negligible. 

The full corner plot is shown in Fig.~\ref{fig_rho_tsz_cib_corner} along with the best-fit values and uncertainties quoted in the 1d posteriors along the diagonal. 
We do not find the LC combination $(X, Y)=$ (MV, \mvcrosscibfree), shown in gray, to be highly constraining and hence do not add those results in Fig.~\ref{fig_rho_tsz_cib}. 
Although the green contours appear tighter in Fig.~\ref{fig_rho_tsz_cib_corner}, the values of $\rhotszcrosscib$ in adjacent bins are more strongly correlated than in the case of the yellow contours. This is intuitively reasonable, since the \cibfree{} estimates enter into both $X$ and $Y$ in that case. 

\subsection{Impact of cluster masking on the tSZ power spectrum}
\label{sec_tszpowspec_cluster_masking}
\begin{figure*}
\centering
\includegraphics[width=0.9\textwidth, keepaspectratio]{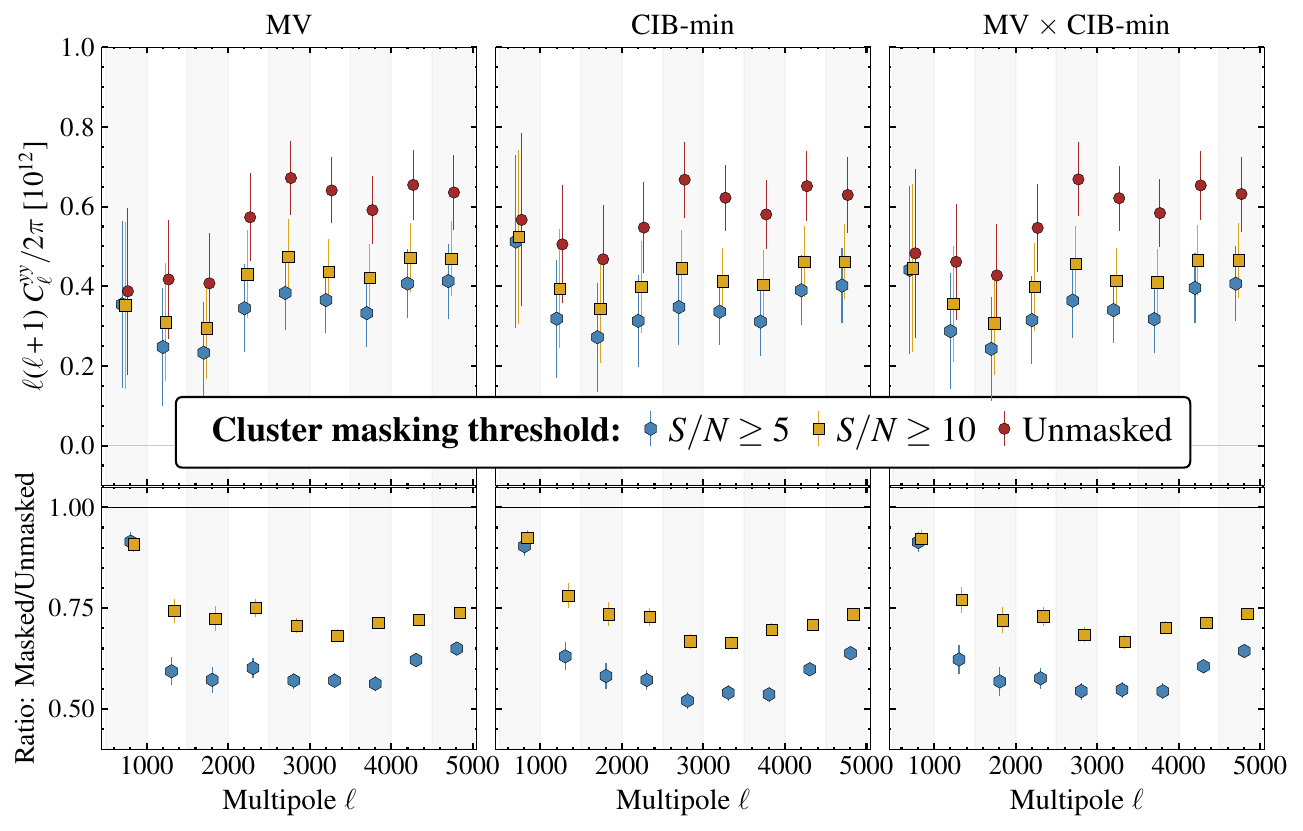}
\caption{Changes to the measured tSZ power spectrum when clusters are masked in the analysis. The top panels show the actual values while the bottom panels show the ratio for different LC schemes. 
Red circles correspond to the baseline unmasked case while the blue hexagons (yellow squares) are after masking clusters detected with $\snr \ge 5$ ($\snr \ge 10$) from the SPT cluster catalog \citep{kornoelje25}. 
The tSZ power reduces by $\times1.8$ and $\times1.5$ at $\ell = 3000$ for the two masking schemes. 
%The ratio of the masked over the unmasked power shows a tilt when comparing large ($\ell<2500$) to small scales ($\ell \ge 2500$) but remain roughly constant at small scales given the measurement errors.
}
\label{ymap_spectra_data_cluster_masking}
\end{figure*}

Several studies have utilized both observational data \citep{george15, hernandez23} and simulations \citep[e.g.,][]{holder07, mccarthy14, raghunathan22b, hadzhiyska23} to investigate how the tSZ power spectrum relates to cluster mass, redshift, and environment, with the ultimate goal of accurately modeling the gas physics of the intracluster medium (ICM).
%The self-similar evolution of clusters suggests that the tSZ signal scales roughly as $Y_{\rm SZ} \propto M^{5/3}$ \citep[][for a recent work]{hadzhiyska23}, and so we can expect the tSZ power spectrum to be dominated by the massive clusters in the survey \citep{mccarthy14}. 

In this section, we quantify the changes in the tSZ power spectrum resulting from masking the detected clusters in the maps. 
To this end, we mask the locations of 285 (63) clusters detected at $\snr \ge 5$ ($\snr \ge 10$) from the \citet{kornoelje25} catalog. 
These roughly correspond to clusters with $\mvir \sim 1.5\ (2.5) \munits$ at a median redshift $z \approx 0.7$ \citep{raghunathan22b, kornoelje25}. 
Similar to source masking, the cluster masking radius also changes based on the detection significance and is roughly between $3^{\prime}-7^{\prime}$.
The results are shown in Fig.~\ref{ymap_spectra_data_cluster_masking}: baseline results where the clusters remain unmasked are shown as red circles, while the blue hexagons (yellow squares) correspond to the results after removing the contribution from clusters with $\snr \ge 5$ ($\snr \ge 10$). 
The top panels show the power spectrum for masked and unmasked cases, and in the bottom panels we present the ratio of masked over the unmasked data points. 
The two masking schemes result in a suppression of the tSZ power spectrum by $\times1.8$ and $\times1.5$ at $\ell = 3000$. 
The power suppression due to cluster masking is roughly constant at $\ell > 1000$. 
However, in the first bin, the power suppression due to cluster masking is much smaller ($\lesssim 10\%$), due to large-scale SPT filtering. 
The decrease in power at small scales is slightly lower than those reported by \citet{raghunathan22b} who predicted roughly $\times2$ reduction of the power (after accounting for the differences in the noise levels between the two studies). 
This discrepancy may stem from differences between the true cluster profiles and the generalized Navarro–Frenk–White (gNFW; \citealt{navarro96}, \citealt{nagai07}) profile assumed in \citet{raghunathan22b}, as well as from large‑scale SPT filtering and astrophysical feedback effects that are not accounted for in that analysis.
We do not attempt to decompose the tSZ power from multiple cluster mass and redshift bins \citep[as done by][]{hernandez23} and leave such an analysis, both with the tSZ-selected clusters and the external catalogs from optical/X-ray observations, to future work. 
This can also help us understand the changes to the tilt of the spectra for different levels of cluster-masking which can give insights into the mass/redshift evolution of the gas physics of clusters. 

% Besides the suppression of the amplitude of the tSZ power spectrum at $\ell = 3000$, studying the changes to the tilt of the spectra for different levels of cluster-masking gives insights into the mass/redshift evolution of the gas physics of clusters. 
% For example, this can probe the mass and redshift dependence of the astrophysical feedback mechanisms. 
% %As indicated by the bottom panels of Fig.~\ref{ymap_spectra_data_cluster_masking}, while we do see a trend moving from large ($\ell<2500$) to small scales ($\ell \ge 2500$), the ratio of the unmasked over the masked spectra remains roughly constant at $\ell \ge 2500$, given the measurement errors. Probing this trend will be a excellent target for the upcoming tSZ power spectrum data from the full \sptthreeg{} \citep{benson14, bender18, sobrin22} or the future Simons Observatory (\simonsobs, \citealt{abitbol25}). 
% As indicated by the bottom panels of Fig.~\ref{ymap_spectra_data_cluster_masking}, while we do see a trend moving from large to small scales, measurement errors prevent us from drawing definitive conclusions. 
% Upcoming data releases from the full \sptthreeg{} \citep{benson14, bender18, sobrin22} dataset and the future Simons Observatory (\simonsobs, \citealt{abitbol25}), will enable more precise measurements of this ratio, and enable further studies tying this possible evolution to cluster astrophysics and their evolution across mass and redshift.

\subsection{Effects of source masking}
\label{sec_tszpowspec_source_masking}
\begin{figure*}
\centering
\includegraphics[width=0.9\textwidth, keepaspectratio]{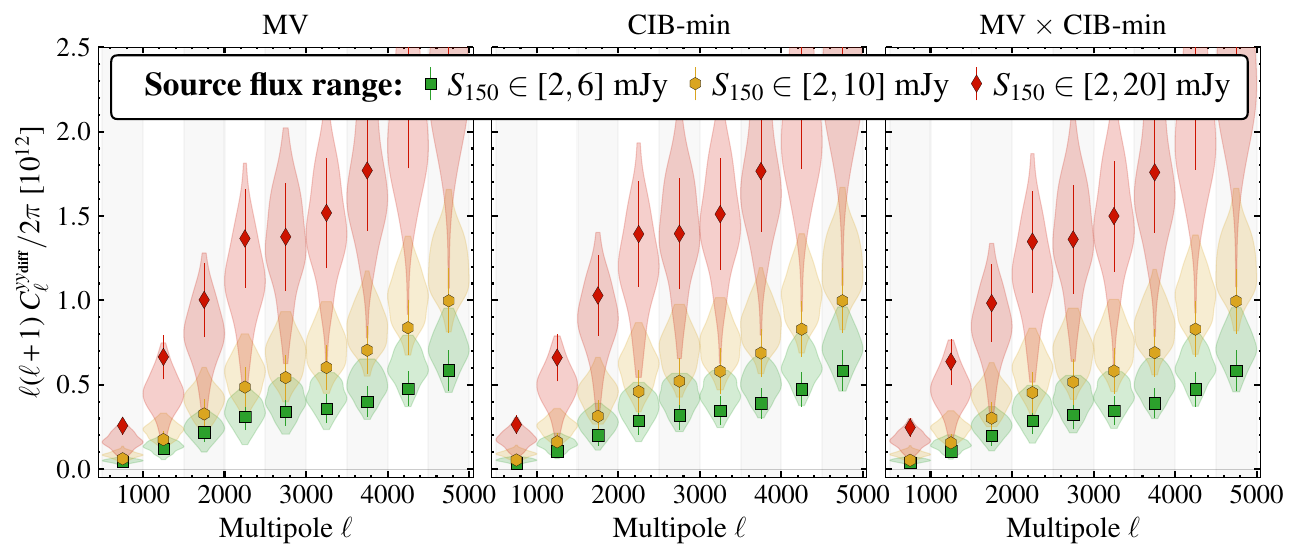}
\caption{Power spectra of the difference maps $C_{\ell}^{yy_{\rm {\bf diff}}}$ computed by subtracting the baseline case of $S_{150} = \baselinemaskingthreshold\ \mjy$ from the maps made with other source masking schemes: $S_{150} = 6\ \mjy$ (green), $S_{150} = 10\ \mjy$ (yellow), and $S_{150} = 20\ \mjy$ (red). 
The violins show the scatter from simulations which include both the statistical and the systematic errors from radio sources. 
The simulations used are CMB + kSZ + tSZ + \texttt{Uncorr-}CIB + \texttt{Uncorr-}radio, and does not contain correlation between tSZ and bright sources.
The observed differences are consistent within the errors, which increase significantly with the masking threshold due to the residual radio power, and we do not observe a significant correlation between tSZ and bright point sources detected in our survey footprint.}
\label{ymap_spectra_data_source_masking}
\end{figure*}

Now we turn to the impact of source masking on the tSZ power spectrum. 
As a reminder, in our baseline analysis, we have removed the contribution of sources with flux $S_{150} \ge \baselinemaskingthreshold\ \mjy$. 
Here, we repeat the analysis after changing the masking threshold to the following: $S_{150} \in [6, 10, 20]\ \mjy$. 
Although we do not differentiate between radio (synchrotron) and dusty sources, this exercise should primarily probe the correlation between the tSZ and bright ($S_{150} \in  [\baselinemaskingthreshold, 20] \ \mjy$) radio galaxies. 
This is because the brightest unmasked radio sources dominate the radio Poisson power in  typical CMB observation bands (\firstsptband/150 GHz), while the CIB power primarily originates from faint sources below the masking threshold \citepalias{reichardt21}. 

We rerun the entire pipeline on both data and simulations (CMB + kSZ + tSZ + \texttt{Uncorr-}CIB + \texttt{Uncorr-}radio) to reconstruct the tSZ map for different ($S_{150} \in [6, 10, 20]\ \mjy$) point source masking schemes. 
Next, we subtract the baseline tSZ map ($S_{150} = \baselinemaskingthreshold\ \mjy$  masking) from each of the above maps and compute the power spectrum of this difference map which we represent as $C_{\ell}^{yy_{\rm {\bf diff}}}$. 
The signals in the unmasked regions are removed in the difference maps, so we are only sensitive to signals within the point source mask regions. 
This contribution is dominated by the signal from unmasked point sources, along with non‑zero contributions from the CMB, kSZ, and tSZ signals at the source locations, as well as correlations among these components. 

We present the results in Fig.~\ref{ymap_spectra_data_source_masking}: green squares, yellow hexagons, and red diamonds correspond to thresholds of $S_{150} \in [6, 10, 20]\ \mjy$ used in the difference maps. 
A first observation is that the signal in the difference maps $C_{\ell}^{yy_{\rm {\bf diff}}}$ is substantially larger than in the baseline‑masking case shown in Fig.~\ref{fig_ymap_spectra_data_plus_literature}, particularly for the 10 and 20 mJy masking thresholds, and is dominated by point sources.
This illustrates the rationale behind our choice of applying a masking threshold of 
$S_{150} = \baselinemaskingthreshold\ \mjy$ for the baseline case. 
The violins in the figure show the scatter in the simulations which includes both the statistical errors and also the systematic error from the uncorrelated radio sources as described in \S\ref{sec_sys_errors}. 
%\refresponse{
Replacing the simulations with analytic estimates of the radio source power using \texttt{de Zotti}, \texttt{Tucci}, and \texttt{Lagache} source counts does not show any difference in the observed scatter. %}
We quantify the agreement between the data and the simulations (which do not include the correlation between the tSZ signal and bright point sources) by computing the $\chi^{2}_{\rm null}$ for each case. We define the $p$-value as the fraction of simulations with $\chi^{2}_{\rm null}$ larger than that of the data, and we find that these values lie in the range 0.1–0.9 across all cases. 
Thus the observed changes are fully consistent with the scatter expected from the simulations, indicating no significant correlations between tSZ and the bright radio sources in our field. 

%%%%%%%%%%%%%%%%%%%%%%%%%%%%%%%%
\section{Conclusion}
\label{sec_conclusion}
In this work, we presented the full shape of the tSZ power spectrum in the range $\ell \in [\lminforps, \lmaxforps]$ reconstructed by combining data from the SPT and \herschel-SPIRE datasets using linear combination techniques. 
%The tSZ maps produced in this 
After removing the expected contribution from undesired residual signals, namely the CMB, CIB, kSZ and radio, we have measured the tSZ power spectrum at $\tszpowerspectrumsnr$.  
We presented numerous checks and quantified the systematic errors due to the assumptions about the residual signals. 
The results are consistent with \planck{} measurements \citep{tanimura22} on the overlapping scales $\ell \in [500, 1100]$, as well as with the recent results from \citet{efstathiou25} and \sptthreeg{} \citep{chaubal26} over a wide range of scales.

We used the difference in the tSZ power spectrum estimates from different combinations of the LC methods to derive the scale-dependent cross-correlation between tSZ and CIB $\rhotszcrosscib$. 
We found the correlation coefficient to be $\sim 0.2$ on large scales but the correlation drops towards zero at $\ell > 2500$. 
The evidence for a non-zero $\rhotszcrosscib$ is at the $\snrofrhotszcibformvandcibfree\sigma$ level. 

We also studied the impact of cluster and source masking on the tSZ power spectrum. 
Masking the detected clusters results in a suppression of the tSZ power, matching with the expectations. 
When we masked clusters detected with $\snr \ge 5$ \citep{kornoelje25}, we found that the tSZ power at $\ell = 3000$ reduced by $\times 1.8$. 
The ratio remains roughly constant at $\ell > 1000$ although the power suppression due to cluster masking in the first bin is much smaller ($\lesssim 10\%$), due to large-scale filtering in SPT maps. 
%By comparing the ratio of the unmasked tSZ power to the masked case, we observed a tilt on large scales $\ell < 2500$ but the power suppression is roughly constant at $\ell \ge 2500$. 
On the other hand, the tSZ power spectrum does not change significantly with different source masking thresholds implying weak tSZ $\times$ radio correlation for the bright $S_{150} \ge 2\ \mjy$ sources in our survey. 
%\refresponse{
In this work, we have not assessed the impact of radio sources below our detection threshold, which may be important for future studies. In principle, any correlation between the tSZ signal and faint radio sources could introduce a bias in the inferred tSZ power spectrum and therefore warrants further investigation.
Additionally, we note the possibility of a radio SZ signal, arising from the Comptonization of the cosmological radio background, as proposed in recent work \citep{holder21, lee22_radiosz}.
Within SPT, we have a dedicated using data from the MeerKAT survey (PID: SCI-20230907-JV-01; PI: Joaquin Vieira; see \citealt{magolego26} for a recent work.) to investigate the tSZ$\times$radio correlation by cross‑correlating the SPT maps with low‑frequency MeerKAT observations. %}

To our knowledge, the maps produced in this work represent the deepest tSZ maps ever produced. 
The analysis pipeline and the tools used in this work are also currently being applied to other datasets within the \sptthreeg{} collaboration to produce deep tSZ maps over much larger footprints covered by the \sptthreeg{} survey \citep{prabhu24}. 

The tSZ maps and the power spectrum, particularly at small scales, as well as the characterization of \tszcrosscib, have important implications for both cosmology and astrophysics. 
Some applications include: 
(a) probing astrophysical feedback processes, which may, in turn, help in shedding light on the origin of the $\seight$ tension \citep{preston23}; 
(b) understanding the gas physics necessary to properly model the ICM pressure profiles \citep{battaglia12}; 
(c) constraining the kSZ power spectrum through priors on tSZ power spectrum amplitude and the shape of the \tszcrosscib{} (\citealt{george15}, \citetalias{reichardt21}); 
(d) calibrating tSZ and multi‑tracer cosmological simulations \citep{stein20, osato23, omori24, lau25, yang25}, which are necessary for analyzing current and next‑generation datasets; and
(e) constraining structure formation and probing the diffuse hot gas in large-scale structures using cross-correlations with lensing and other tracers of the dark matter field.
As a result, the tSZ power spectrum measurements and the maps presented in this work have a wide variety of applications, and this work also lays the groundwork for studies from upcoming data releases from \sptthreeg{} \citep{prabhu24} and Simons Observatory \citep{abitbol25}. 

%%%%%%%%%%%%%%%%%%%%%%%%%%%%%%%%%%%%%%%%%%%%%%%%%%%%

\section*{Data products and availability}
We release the tSZ maps and bandpowers along with the associated products and plotting scripts used in this work. 
They can be downloaded from this:
\begin{itemize}
    \item{{\texttt{gitrepo}$^{\text{\faGithub}}$}: \url{https://github.com/sriniraghunathan/tSZ_2pt_SPT_SPIRE}.}
    \item{\texttt{SPT\_data\_link}: \url{https://pole.uchicago.edu/public/data/tsz_2pt_raghunathan26/index.html}.}
    %\item{\refresponse{\dataset[\texttt{Zenodo}: https://doi.org/10.5281/zenodo.19226440]{https://doi.org/10.5281/zenodo.19226440}}
    \item{\dataset[\texttt{Zenodo}: https://doi.org/10.5281/zenodo.19226440]{https://doi.org/10.5281/zenodo.19226440}
    
}

\end{itemize}

%%%%%%%%%%%%%%%%%%%%%%%%%%%%%%%%%%%%%%%%%%%%%%%%%%%%
\section*{Acknowledgments}

SR dedicates this work to the loving memory of Eric Baxter ---esteemed colleague, friend, and mentor--- whose significant contributions to the study of CMB secondaries and cross-correlations were invaluable.

We thank Ian McCarthy for providing the \flamingo{} power spectrum curves for various astrophysical feedback and cosmological models. 
%\refresponse{
We thank the anonymous reviewer for useful feedback that helped in shaping this manuscript better. %}

SR acknowledges support by the Illinois Survey Science Fellowship from the Center for AstroPhysical Surveys at the National Center for Supercomputing Applications; support of Michael and Ester Vaida, and the National Science Foundation via award OPP-1852617; and also the support from Universities Research Association’s Visiting Scholars Program fellowship. 

This work made use of the following computing resources: Illinois Campus Cluster, a computing resource that is operated by the Illinois Campus Cluster Program (ICCP) in conjunction with the National Center for Supercomputing Applications (NCSA) and which is supported by funds from the University of Illinois Urbana-Champaign; the computational and storage services associated with the Hoffman2 Shared Cluster provided by UCLA Institute for Digital Research and Education's Research Technology Group; OSG Consortium \citep{osg06, osg07, pordes07, osg09}, which is supported by the National Science Foundation awards \#2030508 and \#2323298; and the computing resources provided on Crossover, a high-performance computing cluster operated by the Laboratory Computing Resource Center at Argonne National Laboratory.

% main SPT
The South Pole Telescope program is supported by the National Science Foundation (NSF) through awards OPP-1852617 and OPP-2332483. Partial support is also provided by the Kavli Institute of Cosmological Physics at the University of Chicago. 
% Argonne
Argonne National Laboratory’s work was supported by the U.S. Department of Energy, Office of High Energy Physics, under contract DE-AC02-06CH11357. 
% Davis
The UC Davis group acknowledges support from Michael and Ester Vaida. 
% Fermilab
Work at the Fermi National Accelerator Laboratory (Fermilab), a U.S. Department of Energy, Office of Science, Office of High Energy Physics HEP User Facility, is managed by Fermi Forward Discovery Group, LLC, acting under Contract No. 89243024CSC000002.
% Melbourne
The Melbourne authors acknowledge support from the Australian Research Council’s Discovery Project scheme (No. DP210102386). 
% Paris
The Paris group has received funding from the European Research Council (ERC) under the European Union’s Horizon 2020 research and innovation program (grant agreement No 101001897), and funding from the Centre National d’Etudes Spatiales. 
% SLAC
The SLAC group is supported in part by the Department of Energy at SLAC National Accelerator Laboratory, under contract DE-AC02-76SF00515.
%%%%%%%%%%%%%%%%%%%%%%%%%%%%%%%%%%%%%%%%%%%%%%%%%%%%
\appendix
\restartappendixnumbering

\section{Systematic uncertainties}
\label{sec_sys_errors}

In this section, we describe our techniques for estimating systematic errors due to incorrect estimation of the expectation spectra, calibration errors, and imperfect knowledge of the experimental beam.

\subsection{Systematic uncertainty in the expectation spectra}
\label{sec_sys_errors_in_exp_spectra}
Misestimation of the undesired components will lead to a bias in the final tSZ estimate. In this section, we perform a set of systematic checks, using analytical and simulation-based methods, to ensure that the tSZ-only power spectra we report are robust against the systematics from these contaminating signals. 
We show the expected contribution from each of these components as colored bands in the top panels of Fig.~\ref{ymap_spectra_data_simulations_sys_checks_plus_errors}: CMB in \cmbresidualcolour, CIB in \cibresidualcolour, kSZ in \kszresidualcolour{} and radio in \radioresidualcolour. 
The bands correspond to the range of possible values for each component and indicate the systematic error associated with the misestimation of each of them. 
We obtain the residuals and the systematic error bands as follows.

\subsubsection{CMB residuals} 
We compute the CMB residual using CMB-only simulations of the $\lcdm$ power spectrum. 
%As shown by the \cmbresidualcolour{} curve in Fig.~\ref{ymap_spectra_data_simulations_sys_checks_plus_errors}, the CMB residuals peak ($\dlcmbres = 0.5 \times 10^{-12}$) around $\ell \sim 3000$, roughly near the peak of the expected tSZ power spectrum \citep[][for example]{mccarthy14}, although we note that the exact amplitude and the shape of the tSZ power spectrum is unknown at these scales. 
As shown by the \cmbresidualcolour{} curve in Fig.~\ref{ymap_spectra_data_simulations_sys_checks_plus_errors}, the CMB residuals peak around $\ell \sim 3000$, with an amplitude of $\dlcmbres \approx 0.5 \times 10^{-12}$. 
This peak is roughly near the peak of the expected tSZ power spectrum \citep[][for example]{mccarthy14}, although we note that the exact amplitude and the shape of the tSZ power spectrum is unknown.
The residuals are nearly the same for all of the spectra combinations shown in Fig.~\ref{ymap_spectra_data_simulations_sys_checks_plus_errors}.

To calculate the systematic errors due to misestimation of the CMB residual, we use the \sptthreeg{} (from the 2018 season) and \planck{} chains \citep{balkenhol23} to sample the cosmological parameters randomly. 
Besides the $\lcdm$ parameters, the chains also contain the $T_{\rm Cal}$ factors for \sptthreeg's \firstsptband, 150, and 220 GHz bands, which we also include. 
We assume an absolute calibration error of 0.5\% and 5\% respectively for the \sptpol{} \citep{henning18, chou25} and \herschel-SPIRE \citep{viero19} bands. 
For each sample, $i$, we obtain the CMB power spectrum using \camb{} software \citep{lewis00} and estimate the corresponding CMB residual in the total measured power spectrum as given in Eq.(\ref{eq_cmb_ilc_residuals}) \citepalias{raghunathan23}
\begin{eqnarray}
C_{\ell_{\rm ILC}}^{{\rm CMB}, i} & = & w_{\ell_{A}} \clcov^{{\rm CMB, i}}\ w_{\ell_{B}}^{\dagger}
\label{eq_cmb_ilc_residuals}
\end{eqnarray}
where $\clcov^{{\rm CMB}, i}$ is the ${\rm N}_{\rm band} \times {\rm N}_{\rm band}$ matrix containing the CMB power spectra between different frequency bands for the $i^{th}$ sample and the frequency-dependent LC weights corresponding to $A, B \in$ [MV, \cibfree] are $w_{\ell}$. 
This approach, which uses the sampled power spectra to compute residuals rather than a full map-based simulation, ignores the sample variance in the systematic errors. However, this is expected to have no impact given how small the systematic errors are. 
We also note that the results of \citet{balkenhol23} constrained cosmology using information only up to 
$\ell = 3000$. 
Consequently, any impact on the CMB spectra arising from differences in cosmological parameters at smaller scales will not be taken into account here. 
However, since the systematic error from the CMB is sub-dominant relative to the other signals at these scales, it should have no impact on our results. 
%The systematic error in this estimate corresponding to the \cmbresidualcolour{} band, however, is much smaller compared to the tSZ signal as evident from the CMB-systematic error in the figure.

\subsubsection{CIB residuals} 
\label{sec_cib_residuals}
Next, we compute the CIB residuals and the associated systematics using \agora{} simulations. 
%The CIB residuals peak ($\dlcibres = 0.5 \times 10^{-12}$) around $\ell \sim 2000$ for both MV and \cibfree{} cases but the residuals reduce significantly to $\dlcibres \le 0.1 \times 10^{-12}$ at $\ell \ge 2500$, which is much smaller than the expected level \citep{mccarthy14} of tSZ. 
The CIB residuals peak around $\ell \sim 2000$, with an amplitude of $\dlcibres \approx 0.5 \times 10^{-12}$, for both the MV and \cibfree{} cases, but the residuals are reduced significantly to $\dlcibres \le 0.1 \times 10^{-12}$ at $\ell \ge 2500$, which is much smaller than the expected level of tSZ.
This is due to the low-noise 220 GHz data from \sptthreeg{} and the addition of the high frequency \herschel-SPIRE data.
While the residual level is generally smaller for the \cibfree{} spectra compared to the MV as expected, the residual levels become comparable for the two maps at $\ell \ge 2000$. 
Since CIB dominates the small scale power, the LC weights for MV are tuned to suppress CIB efficiently, resulting in very low residuals, similar to the \cibfree{} case. 
%In fact, the LC weights between the MV and \cibfree{} cases change by $\le 10\%$ for SPT bands at $\ell \ge 2000$. While the differences are larger for \herschel-SPIRE channels, the LC weights for \herschel-SPIRE bands are typically 2-4 orders of magnitude smaller compared to the 150 GHz band at all scales since these bands are primarily being used for CIB mitigation (which is much brighter at the high frequencies compared to the SPT bands).

To get the systematic error band, we take the CIB-only portion of the \agora{} simulations and multiply the map in each SPT band independently with a factor sampled from a normal distribution, $\mathcal{N}(1, \sigma^{2})$, with $\sigma = \sqrt{\cibtweakingsigma}$. 
To be more specific, the scaling modifies the auto- and cross-power spectra of the CIB in each frequency band independently, which introduces decorrelation in the CIB between different bands.
Since the CIB power has been measured at very high significance ($\sim25-35\sigma$), particularly on small-scales, by both \act{} \citep{louis25} and SPT \citepalias{reichardt21}, the 20\% variance used for this exercise is a conservative estimate.
This random scaling of the maps introduces a scatter in the CIB SED for each simulation. Since the LC weights are fixed, this leads to an estimation of the CIB residual in each simulation that is different from our baseline setup. 
We perform this scaling \howmanycibtweaks{} times, and the \cibresidualcolour{} band in Fig.~\ref{ymap_spectra_data_simulations_sys_checks_plus_errors} shows the resulting spread in the CIB residuals. 
We note that this process gives a higher level of residual compared to scaling all the bands up or down by the same factor. 
Note that, in the above procedure, we have not scaled the CIB power in \herschel-SPIRE bands, since the CIB power has been measured with extremely high $\snr$ in those bands, and the simulations we use are calibrated to match the CIB power measured by \herschel-SPIRE.
If we also tweak the CIB power in \herschel-SPIRE bands, then the width of the band increases by $\sim15-20\%$ at all scales. 
Similarly, if we introduce a 30\% scatter, we observe that the width of the band increases by $\sim45\%$. 
Finally, for this exercise, we have ignored the sample variance in CIB and have only tweaked a single \agora{} CIB realization since we find that the scatter between different realizations is negligible, $\times2.5$ smaller, compared to the scatter because of the above scaling.

\subsubsection{Radio residuals} We follow the same setup as described above for the CIB and assume a 20\% scatter to quantify the radio residuals and the associated systematics.
This assumption is conservative, given the measurement errors in the radio power reported by (\citealt{george15}, \citetalias{reichardt21}), and \citet{louis25}. 
In addition, we have examined the radio power under modifications to the underlying source counts (see below) and find that the resulting change is much smaller than the assumed scatter. 
The residuals are obtained using the radio-only portion of \agora{}, and the systematic band, shown as \radioresidualcolour{} in Fig.{\ref{ymap_spectra_data_simulations_sys_checks_plus_errors}}, is obtained by scaling the maps from individual frequency bands. 
The radio residuals tend to increase with multipole with $\dlradres = 0.1\ (0.35)  \times 10^{-12}$ at $\ell = 3000\ (5000)$ for all the three panels in the figure. 
%Note that even though the spectra of radio sources 

For radio residuals, in addition to using the \agora{} simulations, we also perform an additional analytical test by modifying the underlying source counts, $dN/dS$. 
Note that the radio source counts in the \agora{} simulations are shown to match the results from \citet{lagache19}.
%We use the prescription in \S 3.2 of \citetalias{raghunathan23} and calculate the expected radio residuals as given in Eq.(\ref{eq_radio_point_source_power}) for three different source count distributions: \texttt{de Zotti} \citep{dezotti05}, \texttt{Tucci} \citep{tucci11}, and \texttt{Lagache} \citep{lagache19}. 
We use Eq.(\ref{eq_cmb_ilc_residuals}) based on the prescription in \S 3.2 of \citetalias{raghunathan23} and calculate the expected radio residuals. 
The $C_{\ell}^{\rm CMB}$ in Eq.(\ref{eq_cmb_ilc_residuals}) is replaced by $C_{\ell}^{\rm Radio}$ as given in Eq.(\ref{eq_radio_point_source_power}) for three different $dN/dS$ distributions: \texttt{de Zotti} \citep{dezotti05}, \texttt{Tucci} \citep{tucci11}, and \texttt{Lagache} \citep{lagache19} as given below:
\begin{widetext} \begin{align}
C_{\ell_{{\nu_{1} \nu_{2}}}}^{\rm Radio} =  \bigintsss_{\alpharad^{\rm min}}^{\alpharad^{\rm max}} d\alpha \bigintsss_{0}^{S_{150}^{\rm max}} dS_{150}\ S_{150}^{2} \frac{dN}{dS_{150}}\ \left( \frac{\nu_{0}^{2}}{\nu_{1} \nu_{2}} \right)^\alpha \mathcal{N} \left[ \alpharad | \bar{\alpha}_{\rm rad}, \alpharadsigma \right],
\label{eq_radio_point_source_power} 
\end{align}\end{widetext}
with flux threshold $\nu_{0} = 150\ {\rm GHz}$, $S_{150}^{\rm max} = \baselinemaskingthreshold\ {\rm mJy}$, and spectral index $\alpharad = -0.76$ with a conservative scatter of $\alpharadsigma = 0.6$ (\citealt{everett20}, \citetalias{reichardt21}). 
We do not show these results in Fig.~\ref{ymap_spectra_data_simulations_sys_checks_plus_errors} to avoid cluttering. 
As mentioned above, the source number counts of radio galaxies in \agora{} matches the results from \citet{lagache19} to within $10\%$ \citep{omori24}, and we also find that the results for \texttt{Lagache} is within the \radioresidualcolour{} band of Fig.~\ref{ymap_spectra_data_simulations_sys_checks_plus_errors}. 
The residuals are slightly lower for \texttt{de Zotti} and \texttt{Tucci} with the latter giving the lowest radio residual. 
However, the difference in the radio residuals between the three source count distributions is much smaller than the \radioresidualcolour{} band obtained using the scalings above. 
The difference in the radio residuals from the three source count distributions compared to the width of the \radioresidualcolour{} band is $\le 5\%$ at $\ell = 3000$ and increases to $\sim50\%$ at $\ell = 5000$.

\subsubsection{kSZ residuals} The baseline kSZ residuals are obtained using \agora{} simulations with a kSZ power that is roughly flat in $D_{\ell}$ with an amplitude $D_{\ell}^{\rm kSZ} = 3\ \uk^{2}$. 
The kSZ residuals are nearly the same for all the spectra and roughly correspond to $\dlkszres = 0.05\ (0.2) \times 10^{-12}$ at $\ell = 3000\ (5000)$.

However, since the kSZ power spectrum has not been detected at high significance, we calculate the systematic from kSZ mismodeling analytically using the same procedure as the one for the CMB described above in Eq.(\ref{eq_cmb_ilc_residuals}).
To this end, we replace the above kSZ amplitude with a uniform distribution $D_{\ell}^{\rm kSZ} \in \mathcal{U}(0, 4)\ \uk^{2}$. 
We repeat this exercise \howmanycibtweaks{} times and obtain the \kszresidualcolour{} band in Fig.~\ref{ymap_spectra_data_simulations_sys_checks_plus_errors}. 

\subsection{Calibration errors}
\label{sec_calib_errors}
To take into account the effect of uncertainties in the absolute calibration, we randomly multiply the maps from all the bands in the simulation with a factor sampled from a normal distribution, $\mathcal{N}(1, \sigma_{\rm Calib}^{2})$. 
The calibration errors were set to be: 0.5\% for SPTpol and \sptthreeg{} maps (i.e., $\sigma_{\rm Calib} = 0.005$ based on \citealt{henning18, balkenhol23, camphuis25}) and 5\% for \herschel-SPIRE maps \citep{swinyard10, viero19}. 
We perform the above operation 100 times and estimate the resulting scatter. 
This scatter, shown using purple color in the bottom panel of Fig.~\ref{ymap_spectra_data_simulations_sys_checks_plus_errors}, is taken into account in the systematic errors along with the uncertainties from the other astrophysical components described above.

\subsection{Uncertainties in beam estimation}
\label{sec_beam_uncertainities}
In this subsection, we investigate the effect of uncertainties in the experimental beam, $B_{\ell}$, on the measured tSZ power spectra. 

\subsubsection{Beam chromaticity} As described in \S\ref{sec_ilc} and \S\ref{sec_tf_beams}, we deconvolve the CMB beam, $B_{\ell}^{\rm CMB}$, from the maps before the LC step. 
To account for the variation of the beam across frequencies for different foreground signals, we convolve these signals with their respective component-dependent beams in the simulations. 
Before the LC step, similar to data, we also deconvolve the $B_{\ell}^{\rm CMB}$ from the simulations to mimic the data. 
Alternatively, if we convolve all foreground signals with $B_{\ell}^{\rm CMB}$ ignoring SED variations, we observe a scale-dependent trend. 
The impact, however, is small with a mean shift of $\sim0.2\sigma$ and a maximum shift of about $\lesssim0.3-0.4\sigma$ around $\ell \sim 2000$ depending on which combination of LC maps is used in the spectrum. 

\subsubsection{Beam uncertainties} As a reminder, the SPT beams are composite real-space beams constructed by combining dedicated sets of planets and point source observations. 
We use stacks of point sources in regions where detector nonlinearities affect the planet scans (\citealt{dutcher21}, N. Huang et al., in preparation). 
In this subsection, we quantify how uncertainties in the SPT beam estimation $\sigma(B_{\ell})$ propagate into our results. 
These uncertainties arise both from noise-induced scatter and from systematic effects associated with the choice of radius used when combining planet and point source observations to construct the composite beam (N. Huang et al., in preparation).
For this test, we alter the component-dependent beams used in the simulation as $B_{\ell}^{\rm comp} \rightarrow B_{\ell}^{\rm comp} + \sigma(B_{\ell}^{\rm comp})$ while deriving the expectation signal for the undesired components. 
The CMB beam without the additional shift $B_{\ell}^{\rm CMB}$ is deconvolved from the simulations before the LC step. 
Similar to the case above, the final tSZ estimates also show a slight scale-dependent trend. 
The maximum shift is in the first bin which moves up $\sim 0.5\sigma$ but the impact is much smaller for the other bins: $\le 0.25\sigma$ for $\ell \le 3000$ and $\le 0.1\sigma$ for $\ell > 3000$. 
The average shift is $\le 0.15 \sigma$ for all LC estimates.

%%%%%%%%%%%%%%%%%%%%%%%%%%%%%%%%%%%%%%%%%%%%%%%%%%%%
%\bibliography{spt_bib_extra, spt}
\IfFileExists{spt_before_1995.bib}
{
\bibliography{spt_before_1995, spt_1995_to_2000, spt_2000_to_2005, spt_2005_to_2010, spt_2010_to_2015, spt_2015_to_2020, spt_2020_to_2025, spt_2025_and_after, tsz_2pt}{}
}
{\bibliography{../../../BIBTEX/spt_before_1995, ../../../BIBTEX/spt_1995_to_2000, ../../../BIBTEX/spt_2000_to_2005, ../../../BIBTEX/spt_2005_to_2010, ../../../BIBTEX/spt_2010_to_2015, ../../../BIBTEX/spt_2015_to_2020, ../../../BIBTEX/spt_2020_to_2025, ../../../BIBTEX/spt_2025_and_after, tsz_2pt}{}}
\bibliographystyle{aasjournalv7}

\ifdefined\PRformat
\bibliographystyle{apsrev4-1}
\fi
%%%%%%%%%%%%%%%%%%%%%%%%%%%%%%%%%%%%%%%%%%%%%%%%%%%%
\end{document}